\documentclass{article}
\usepackage{color}
\usepackage{graphicx}
\usepackage{amsmath, amsthm}
\usepackage{mathtools}
\usepackage{natbib}
\usepackage{url}
\RequirePackage[colorlinks,citecolor=blue,urlcolor=blue]{hyperref}
\usepackage{xcolor}
\usepackage{chngcntr}
\usepackage{subfig}
\usepackage{hyperref}
\usepackage{algorithm}
\usepackage[noend]{algpseudocode}
\usepackage{setspace}
\newtheorem{theorem}{Theorem}

\usepackage{xfrac}
\definecolor{forest}{rgb}{0.133,0.545,0.133}
\usepackage{multirow}
\usepackage{amsfonts}
\newtheorem{lemma}{Lemma}

\usepackage{enumitem}
\usepackage{caption}
\usepackage{array}
\usepackage{makecell}
\numberwithin{equation}{section}

\evensidemargin=\oddsidemargin
\usepackage{etoolbox}

\newif\ifabbreviation
\pretocmd{\thebibliography}{\abbreviationfalse}{}{}
\AtBeginDocument{\abbreviationtrue}

\begin{document}
	\newcommand{\bb}{\boldsymbol{\beta}}

	\title{Computationally Efficient Experimental Design with Generalized Posteriors}


	\author{Luke Hagar\footnote{Luke Hagar is the corresponding author and may be contacted at \url{l.hagar@uq.edu.au}.} \hspace{35pt} James M. McGree$^{\dagger}$ \bigskip \\ 
 $^*$\textit{Clinical Trials Capability, The University of Queensland} \\
$^{\dagger}$\textit{School of Mathematical Sciences, Queensland University of Technology}}

	\date{}

	\maketitle

	\begin{abstract}

The hybrid approach to experimental design aims to control frequentist operating characteristics of Bayesian decision procedures. These operating characteristics are assessed by simulating sampling distributions of posterior summaries under assumed data-generation processes that also define posterior distributions. Model misspecification can distort effect estimation and compromise control over operating characteristics. Generalized posterior distributions are defined using loss functions that characterize data generation under fewer assumptions, enhancing the robustness of Bayesian analysis and study design. However, computationally efficient design methodology with generalized posteriors is lacking. We propose an efficient method to determine suitable sample sizes and decision criteria associated with generalized posteriors under the hybrid approach. Using theoretical results to model posterior summaries as linear functions of the sample size, we efficiently assess operating characteristics throughout the sample size space given simulations conducted at only two sample sizes. While the benefits of the proposed methodology are illustrated by redesigning an adaptive clinical trial with time-to-event outcomes, we overview our framework’s broader applicability to experiments involving Bayesian analogues to M-estimation. 

		\bigskip

	\end{abstract}

	\maketitle

	\baselineskip=19.5pt



\section{Introduction}\label{sec:intro}

 In scientific experiments, Bayesian decision procedures provide several advantages, including the formal incorporation of prior information 
 and a principled framework for data-driven adaptation in study design. Despite these advantages, standard Bayesian methods are constrained by the need to fully specify parametric models via likelihood functions \citep{bernardo2009bayesian}. Even if the target of inference for a study is low dimensional, one must specify a (potentially high-dimensional) model for data generation. The misspecification of such models can compromise the validity of Bayesian decision procedures \citep{kleijn2012bernstein}.  

 Generalized posterior distributions mitigate several limitations associated with parametric likelihood specification: they can enhance robustness to model misspecification and outliers, reduce computational complexity, and facilitate inference without full specification of nuisance parameters. The generalized posterior is proportional to a \color{black} loss function \color{black} times a prior distribution \citep{miller2021asymptotic}.  Generalized posteriors have been defined using various \color{black} loss functions \color{black} \citep{jiang2008gibbs,bissiri2016general}, including those based on composite \citep{pauli2011bayesian, friel2012bayesian} or partial \citep{raftery1994accounting, sinha2003bayesian} likelihoods. Generalized posteriors have appeared in the literature under various names, such as general belief updates \citep{bissiri2016general}, M-posteriors \citep{marusic2025theoretical}, and quasi-posteriors \citep{chernozhukov2003mcmc}.

 The use of generalized posteriors in experimental design is a developing area. Recently, \citet{overstall2025gibbs} proposed a general framework for optimal design of experiments that employ inference based on generalized posteriors. This framework was applied by \citet{mcgree2025approach} to design adaptive clinical trials with time-to-event outcomes. 
 Their methodology leveraged a partial likelihood to facilitate inference that is robust to the form of the baseline hazard function -- resulting in more accurate estimation of treatment effects, stricter control of type I error rates, and better sample size recommendations.

 This work focuses on studies in which decision making is based on posterior summaries -- various posterior and posterior predictive probabilities -- from generalized posteriors. Even when Bayesian posterior summaries inform decision making, sample sizes and decision criteria are often chosen to control frequentist operating characteristics. Regulatory agencies generally require strict control of operating characteristics in clinical studies \citep{fda2019adaptive, fda2026bayesian}, and the frequentist operating characteristics of Bayesian designs are of broader interest \citep{ye2022normalized,deng2024metric, hagar2024fast}. Approaches that control frequentist operating characteristics associated with Bayesian decision procedures are often called hybrid approaches \citep{berry2010bayesian}. 

 Under the hybrid approach, suitable sample sizes and decision criteria are found by estimating sampling distributions of posterior summaries using simulation \citep{wang2002simulation}. For each sample size considered, many repetitions of an experiment are simulated according to a given data-generation process to estimate  operating characteristics.  \citet{gubbiotti2011bayesian} defined two methodologies for specifying the data-generation process; the predictive approach accommodates uncertainty in this process, whereas the conditional approach does not. With either approach, design procedures based on naive simulation are computationally expensive and ignore information provided by the large-sample theory associated with generalized posteriors (see e.g., \citet{miller2021asymptotic, marusic2025theoretical}). A general and computationally efficient method for design with generalized posteriors under the hybrid approach would make experimentation based on robust Bayesian inference more accessible to practitioners.

 With parametric inference based on a fully specified likelihood, \citet{hagar2025economical} proposed a method to estimate the sampling distribution of posterior probabilities throughout the sample size space using estimates of the sampling distribution at only two sample sizes. That framework has been extended to accommodate Bayesian designs with clustered data \citep{hagar2026design}, sequential analysis and certain types of posterior predictive probabilities \citep{hagar2026group}, and multiple hypothesis testing \citep{hagar2026fdr}. However, the theoretical results that substantiate this efficient framework for study design rely on the standard Bernstein-von Mises (BvM) theorem  \citep{vaart1998bvm}, 
 precluding the use of generalized posteriors. 
 
 In this work, we extend the framework from \citet{hagar2026group} to accommodate (i) experiments based on robust Bayesian inference with generalized posteriors and (ii) a broader class of posterior predictive probabilities used to analyze censored data. These extensions stem from new theoretical results that show the quantiles of the sampling distribution of the logits of posterior summaries depend linearly on the sample size. Based on numerical studies conducted at only two sample sizes, we can thus infer the sampling distribution at new sample sizes. This design framework promotes a computationally efficient and broadly applicable approach to simulation-based design for experiments that leverage Bayesian analogues to M-estimation. 

 The remainder of this article is structured as follows. In Section \ref{sec:methods}, we introduce preliminary concepts regarding generalized posteriors. Section \ref{sec:methods.2} overviews posterior summaries commonly used in adaptive designs and their sampling distributions. In Section \ref{sec:proxy}, we construct a proxy to a joint sampling distribution of posterior and posterior predictive probabilities based on generalized posteriors and prove novel theoretical results about these proxies. We adapt these theoretical results to develop a procedure to select sample sizes and decision criteria that requires estimation of the true sampling distribution at only two sample sizes in Section \ref{sec:power}. In Section \ref{sec:ex}, we illustrate the performance of our methodology in non-adaptive contexts with two stylized examples. Section \ref{sec:orvac} showcases the proposed methods with an example based on a real adaptive clinical trial. We conclude and discuss extensions to this work in Section \ref{sec:disc}.


\section{Generalized Posteriors}\label{sec:methods}

We first introduce generalized posteriors in a non-adaptive context where the sample size $n$ is fixed before collecting data. We consider design for an experiment with an i.i.d.\ sample of size $n$, $\boldsymbol{x}_n = (X_1, \dots, X_n) \in \mathcal{X}^n$, where $\mathcal{X} \in \mathbb{R}^d$ denotes the sample space for a single observation. 
This framework accommodates studies based on scalar observations and those with additional covariates that may comprise a regression model or denote censoring. The data-generation process $F$ for $\boldsymbol{x}_n$ need not belong to a family of parametric distributions. 
A table summarizing the notation used in this paper is provided in Appendix A of the online supplement.

We consider a target of inference $\theta \in \Theta$ that is a functional $W$ of the data-generation process: $\theta = W(F)$. We focus on targets of inference that can be framed within the M-estimation framework. The M-estimator $\hat{\theta}^{_{(n)}}$ is defined as a minimizer of the form
\begin{equation}\label{eq:M.est}
    \hat{\theta}^{_{(n)}} = \arg\min_{\theta\in\Theta}\dfrac{1}{n}\sum_{i=1}^n\rho(\boldsymbol{X}_i, \theta),
\end{equation}
where $\rho: \mathcal{X}\times \Theta \rightarrow \mathbb{R}$ is a loss function that defines both the target of inference and its generalized posterior. We let $\pi(\theta)$ be a prior distribution for $\theta$. The generalized posteriors we use for decision making  are formulated as
\begin{equation}\label{eq:M.post}
\pi(\theta~|~\boldsymbol{x}_n) = \dfrac{\exp(-\omega\sum_{i=1}^n \rho(\boldsymbol{X}_i, \theta))\pi(\theta)}{\int_{\Theta}\exp(-\omega\sum_{i=1}^n \rho(\boldsymbol{X}_i, \theta))\pi(\theta)d\theta},
\end{equation}
where $\omega > 0$ controls the rate of learning about $\theta$ from prior to posterior.

To illustrate the flexibility of this inferential framework, we consider two example loss functions for a two-group experiment with time-to-event data. In this experiment, data $\boldsymbol{X}_i = (Y_i, A_i, \delta_i)$ from subjects $i = 1, \dots, n$ are available, where $Y_i$ is a time-to-event outcome for subject $i$, $A_i \in \{0,1\}$ is the treatment assignment, and $\delta_i \in \{0,1\}$ denotes whether this outcome has been right censored. \color{black} The following example loss functions illustrate that generalized posteriors encompass -- but are not constrained to -- standard parametric Bayesian inference. \color{black}

The first loss function $\rho_1(\boldsymbol{X}_i, \theta, \eta) = -\delta_i\log(\eta) -\delta_i\theta A_i + \eta y_i e^{\theta A_i}$ corresponds to the negative log-likelihood of a proportional hazards (PH) model with an exponential baseline hazard. \color{black}This baseline hazard is specified via a nuisance parameter $\eta$ that can be marginalized out of $\pi(\theta~|~\boldsymbol{x}_n)$. \color{black}The second loss function is the negative log-partial likelihood 
    \begin{equation}\label{eq:partial}
        \rho_2(\boldsymbol{X}_i, \theta) = -\delta_i \log \left[\dfrac{\exp(\theta A_i)}{\sum_{j \in K_i}\exp(\theta A_j)}\right],
    \end{equation}
    where $K_i$ denotes the risk set (i.e., the set of subjects who have not yet experienced the event of interest or been censored) at the time $y_i$, $i = 1, \dots, n$.

    We suppose the target of inference $\theta$ is the log-hazard ratio for both loss functions. The negative log-likelihood that defines $\rho_1$ corresponds to a distribution that is either correctly specified or serves as a proxy model (e.g., if the true baseline hazard for the PH model is not exponential). The loss function $\rho_2$ instead facilitates semi-parametric inference for $\theta$ in a PH model that is robust to the form of the baseline hazard. \color{black}For now, we suppose that $\omega = 1$. This learning rate is suitable for both loss functions in the absence of loss-function misspecification: \color{black} $\rho_1$ corresponds to a complete likelihood, and confidence intervals corresponding to the partial likelihood that informs $\rho_2$ have asymptotically correct coverage \citep{cox1975partial}. \color{black} We discuss the case where $\omega \ne 1$ and implications of loss-function misspecification in Section \ref{sec:ex.2}. \color{black}

    The generalized posterior in (\ref{eq:M.post}) informs decision making for the experiments in this paper. The complementary interval hypotheses that support decision making are $H_0: \theta \notin (\delta_L, \delta_U)$ vs.\ $H_1: \theta \in (\delta_L, \delta_U)$,
where $-\infty \le \delta_L < \delta_U \le \infty$. These hypotheses are defined with respect to general notation for the interval $(\delta_L, \delta_U)$ such that the goal of the experiment (e.g., determining superiority, non-inferiority, or practical equivalence) is achieved when the data support $H_1$. For example, $\delta_L = -\infty$ is a natural choice for comparisons based on efficacy or superiority when smaller values of $\theta$ are preferred. 

\color{black}
While a loss function defines the generalized posterior in (\ref{eq:M.post}), the decision rules based on $\pi(\theta~|~\boldsymbol{x}_n)$ that we consider are not derived by specifying a full decision-theoretic loss function. \color{black}For a non-adaptive design, we instead conclude that $H_1$ is supported when the posterior probability $\Pr(H_1~|~\boldsymbol{x}_n) \ge \gamma$ for some pre-specified success threshold $\gamma \in [0,1]$. Posterior summaries that inform more complex decision procedures in adaptive designs are introduced in Section \ref{sec:methods.2}. 

\section{Joint Sampling Distributions of Posterior Summaries}\label{sec:methods.2}

\subsection{Posterior Summaries in Adaptive Designs}

While our design methods can be applied in non-adaptive settings, this paper emphasizes group sequential experiments with stopping rules based on posterior summaries. Group sequential designs have $T$ potential analyses indexed by $t \in \{1, \dots, T\}$. At the $t^{\text{th}}$ analysis, the cumulatively accrued $n_t$ observations comprise the data $\boldsymbol{x}_{n_t}$. To discuss designs with arbitrary $T$, we refer to each analysis generally by its index $t$, and we note that analyses 1 to $T-1$ are interim analyses. 
Our notation and theoretical development in Section \ref{sec:proxy} are with respect to joint sampling distributions across all $T$ analyses for a hypothetical design without early stopping. Nevertheless, we can use these sampling distributions to design group sequential experiments with interim decision rules under conditions described in Section \ref{sec:power}. 

We accommodate sequential designs with stopping rules based on various posterior or posterior predictive probabilities. We first consider stopping rules based on posterior probabilities. We index the joint sampling distribution of posterior probabilities using the sample size for the first analysis. In particular, we index by a sample size $n$ such that $\{n_t\}_{t=1}^T = n \times \{c_t\}_{t=1}^T$ for some constants $c_1=1$ and $\{c_t\}_{t=2}^T > 1$. The data $\boldsymbol{x}_n = \{\boldsymbol{x}_{n_t} \}_{t=1}^T$ across all analyses define a vector of $T$ posterior probabilities about the hypothesis $H_1$:
  \begin{equation}\label{eq:pp}
           \boldsymbol{\tau}_{\text{PP}}(\boldsymbol{x}_n)  = 
           \begin{bmatrix}
           \tau_{\text{PP}, 1}(\boldsymbol{x}_n) \\
           \vdots \\
           \tau_{\text{PP}, T}(\boldsymbol{x}_n) 
         \end{bmatrix} = \begin{bmatrix}
           \Pr(H_{1}~|~\boldsymbol{x}_{n_1}) \\
           \vdots \\
           \Pr(H_{1}~|~\boldsymbol{x}_{n_T}) 
         \end{bmatrix}.
\end{equation} 
For an adaptive design, we may conclude that $H_1$ is supported by comparing $\boldsymbol{\tau}_{\text{PP}}(\boldsymbol{x}_n)$ to success thresholds $\{\gamma_t\}_{t=1}^T \in [0,1]^T$. An ongoing experiment may be stopped for success at analysis $t$ if $\tau_{\text{PP}, t}(\boldsymbol{x}_n) \ge \gamma_t$. The non-adaptive design discussed in Section \ref{sec:methods} is recovered when $T=1$. 

We next describe stopping rules for adaptive designs based on two types of posterior predictive probabilities. 
The first type of posterior predictive probability we consider is often used in adaptive designs with time-to-event data. In such designs, we may need to make an interim decision at analysis $t < T$ before observing all event or censoring times in $\boldsymbol{x}_{n_t}$. The outcomes for the $\tilde{n}$ units that are still under observation will be ``temporarily'' censored at the time of the $t^{\text{th}}$ analysis less their study enrollment time. To account for the uncertainty arising from these $\tilde{n}$ outcomes, we integrate over truncated event times from the posterior predictive distribution \citep{gelman1996posterior, berry2010bayesian, saville2014utility}, which characterizes the distribution of future data according to the current posterior. We formally define this distribution as 
  \begin{equation}\label{eq:ppd}   
 \dot{F}(\tilde{x}~|~\boldsymbol{x}_{n_t}) =         \int_{\Theta}\tilde{F}(\tilde{x}~|~\theta, \boldsymbol{x}_{n_t}) \pi(\theta~|~\boldsymbol{x}_{n_t})d\theta,
\end{equation}
where $\tilde{F}$ is a data-generation process that may differ from $F$. 

The notation in (\ref{eq:ppd}) emphasizes that $\tilde{F}$ depends on the value for the target of inference $\theta$. However, $\theta$ itself may not be sufficient for data generation in a framework with generalized posteriors. We also note that $\tilde{F}$ depends on $\boldsymbol{x}_{n_t}$ in contexts with truncation; all truncated event times distributed according to $\dot{F}$ must exceed the corresponding temporary censoring times. For example, a generalized posterior may facilitate survival analysis using the loss function in (\ref{eq:partial}) that is based on the partial likelihood. Truncated data for this example are generated from $\dot{F}$ such that the log-hazard ratio is drawn from $\pi(\theta~|~\boldsymbol{x}_{n_t})$, but the baseline hazard is estimated from the data through other means (e.g., a flexible spline method). While $\tilde{F}$ may depend on nuisance parameters like the baseline hazard, we do not formally incorporate such parameters into (\ref{eq:ppd}) to simplify notation. As detailed later, $\dot{F}$ may also characterize the distribution of untruncated observations. In that case, $\tilde{F}(\tilde{x}~|~\theta, \boldsymbol{x}_{n_t})$ reduces to $\tilde{F}(\tilde{x}~|~\theta)$ under the i.i.d.\ assumption for $\boldsymbol{x}_{n}$.  

Integration over the $\tilde{n}$ truncated event times from $\dot{F}$ yields the \emph{interim} predictive probability of concluding success: 
  \begin{equation}\label{eq:ipp.def}   
           \Pr\{\Pr(H_{1}~|~\tilde{\boldsymbol{x}}_{n_t}) \ge \gamma_{t}~|~\boldsymbol{x}_{n_t}\} = \int_{\Theta} \Pr\{\Pr(H_{1}~|~\tilde{\boldsymbol{x}}_{n_t}) \ge \gamma_{t}~|~\theta, \color{black}\boldsymbol{x}_{n_t}\color{black}\}\pi(\theta~|~\boldsymbol{x}_{n_t})d\theta,
\end{equation}
where $\tilde{\boldsymbol{x}}_{n_t}$ represents a sample where the temporarily censored times in $\boldsymbol{x}_{n_t}$ are replaced with truncated event times from $\dot{F}$. The sample $\tilde{\boldsymbol{x}}_{n_t}$ may incorporate further censoring (e.g., maximum follow-up times). The data, indexed by the sample size $n$ for the first analysis, define a vector of $T-1$ interim predictive probabilities:
  \begin{equation}\label{eq:ipp}
           \boldsymbol{\tau}_{\text{IP}}(\boldsymbol{x}_n)
    = \begin{bmatrix}
           \tau_{\text{IP}, 1}(\boldsymbol{x}_n) \\
           \vdots \\
           \tau_{\text{IP}, T-1}(\boldsymbol{x}_n)
         \end{bmatrix} = \begin{bmatrix}
           \Pr\{\Pr(H_{1}~|~\tilde{\boldsymbol{x}}_{n_1}) \ge \gamma_{1}~|~\boldsymbol{x}_{n_1}\} \\
           \vdots \\
           \Pr\{\Pr(H_{1}~|~\tilde{\boldsymbol{x}}_{n_{T-1}}) \ge \color{black}\gamma_{T-1}\color{black}~|~\boldsymbol{x}_{n_{T-1}}\}
         \end{bmatrix}.
\end{equation}
Stopping rules based on posterior predictive probabilities will be discussed in Section \ref{sec:power}. 

Interim predictive probabilities are estimated using intensive simulation. First, a value $\theta_m$ 
is drawn from the generalized posterior $\pi(\theta~|~\boldsymbol{x}_{n_t})$. Second, $\tilde{n}$ truncated event times are generated from $\tilde{F}(\tilde{x}~|~\theta_m, \boldsymbol{x}_{n_t})$; these $\tilde{n}$ times replace the temporarily censored times in $\boldsymbol{x}_{n_t}$ to create $\tilde{\boldsymbol{x}}_{n_t, m}$. Third, the posterior probability $\Pr(H_{1}~|~\tilde{\boldsymbol{x}}_{n_t, m})$ is computed using $\pi(\theta~|~\tilde{\boldsymbol{x}}_{n_t, m})$. These three steps are repeated $M$ times. The interim predictive probability in (\ref{eq:ipp.def}) is estimated as $M^{-1}\sum_{m = 1}^M\mathbb{I}\{\Pr(H_{1}~|~\tilde{\boldsymbol{x}}_{n_t, m}) \ge \gamma_t\}$.

The second type of posterior predictive probability we consider is introduced in the context of time-to-event data, but it can be used with any outcome type. 
The \emph{final} predictive probability of concluding success for an analysis $t < T$ is the probability that $\tau_{\text{PP}, T}(\boldsymbol{x}_n)$ will be at least $\gamma_T$ given the current data $\boldsymbol{x}_{n_t}$. This probability is considered when data generation for (i) the $n_T - n_t$ observations from the upcoming stages and (ii) the $\tilde{n}$ temporarily censored observations is characterized by $\dot{F}$:
\begin{equation}\label{eq:fpp.def}   
           \Pr\{\Pr(H_{1}~|~\tilde{\boldsymbol{x}}_{n_T}) \ge \gamma_{T}~|~\boldsymbol{x}_{n_t}\} = \int_{\Theta} \Pr\{\Pr(H_{1}~|~\tilde{\boldsymbol{x}}_{n_T}) \ge \gamma_{T}~|~\theta, \color{black}\boldsymbol{x}_{n_t}\color{black}\}\pi(\theta~|~\boldsymbol{x}_{n_t})d\theta,
\end{equation}
where $\tilde{\boldsymbol{x}}_{n_T}$ is such that the observed sample is augmented with $n_T - n_t$ observations from $\dot{F}$. In the sample $\tilde{\boldsymbol{x}}_{n_T}$, temporarily censored times may be replaced with truncated event times, but this consideration is not applicable for outcomes that are not time to event. The final predictive probability in (\ref{eq:fpp.def}) can be estimated using a process similar to that described for the interim predictive probability, where $n_T -n_t$ (untruncated) outcomes are simulated from $\tilde{F}(\tilde{x}~|~\theta_m)$ in the second step. The vector of $T-1$ final predictive probabilities for the experiment are
  \begin{equation}\label{eq:fpp}
           \boldsymbol{\tau}_{\text{FP}}(\boldsymbol{x}_n)
    = \begin{bmatrix}
           \tau_{\text{FP}, 1}(\boldsymbol{x}_n) \\
           \vdots \\
           \tau_{\text{FP}, T-1}(\boldsymbol{x}_n)
         \end{bmatrix} = \begin{bmatrix}
           \Pr\{\Pr(H_{1}~|~\tilde{\boldsymbol{x}}_{n_T}) \ge \gamma_{T}~|~\boldsymbol{x}_{n_1}\} \\
           \vdots \\
           \Pr\{\Pr(H_{1}~|~\tilde{\boldsymbol{x}}_{n_T}) \ge \gamma_{T}~|~\boldsymbol{x}_{n_{T-1}}\}
         \end{bmatrix}.
\end{equation} 

We do not consider posterior predictive probabilities at analysis $T$ because all outcomes should be observed, or permanently censored, at the end of the experiment. In Section \ref{sec:power}, we assess design operating characteristics using the joint sampling distribution of posterior probabilities in (\ref{eq:pp}), interim predictive probabilities in (\ref{eq:ipp}), and final predictive probabilities in (\ref{eq:fpp}). We jointly refer to these probabilities as
  \begin{equation*}\label{eq:both}
           \boldsymbol{\tau}(\boldsymbol{x}_n)
    = \begin{bmatrix}
           \boldsymbol{\tau}_{\text{PP}}(\boldsymbol{x}_n) \\
           \boldsymbol{\tau}_{\text{IP}}(\boldsymbol{x}_n) \\
           \boldsymbol{\tau}_{\text{FP}}(\boldsymbol{x}_n)
         \end{bmatrix}.
\end{equation*} 

\subsection{A Computational Bottleneck}

To estimate the sampling distribution of $\boldsymbol{\tau}(\boldsymbol{x}_n)$ via simulation, we consider various data-generation mechanisms. For simulation repetition $r = 1, \dots, R$, data $\boldsymbol{x}_{n,r}$ are generated across all stages of the experiment according to a given data-generation process $F_r$, and $\boldsymbol{\tau}(\boldsymbol{x}_{n,r})$ is computed. A family of data-generation processes $\mathcal{F}$ may be used to accommodate uncertainty in $\{F_r \}_{r=1}^R$; examples are provided in Section \ref{sec:ex}. If uncertainty is incorporated into $\mathcal{F}$, it is common to fix the model used for data generation but allow (a subset of) its parameters to vary across simulation repetitions according to some probability distribution.  This notation accommodates the conditional and predictive approaches, where $\mathcal{F}$ is degenerate under the conditional approach (i.e., all $\{F_r \}_{r=1}^R$ are the same). The collection of obtained $\{\boldsymbol{\tau}(\boldsymbol{x}_{n,r})\}_{r = 1}^R$ values estimates the joint sampling distribution of $\boldsymbol{\tau}(\boldsymbol{x}_n)$.

We discuss how the sample sizes $\{n_t\}_{t=1}^T$ and decision thresholds impact the operating characteristics of adaptive designs in Section \ref{sec:power}. For every value of $n = n_1$ considered, we must obtain a collection of $\{\boldsymbol{\tau}(\boldsymbol{x}_{n,r})\}_{r = 1}^R$ values via simulation to estimate operating characteristics, while tuning the decision thresholds, to find a suitable design. The process to obtain the $\{\boldsymbol{\tau}(\boldsymbol{x}_{n,r})\}_{r = 1}^R$ values is computationally intensive; however, we can reduce the computational burden by using previously estimated sampling distributions of $\boldsymbol{\tau}(\boldsymbol{x}_{n})$ to estimate operating characteristics at new $n$ values. We can use this process to efficiently design experiments based on generalized posteriors. We propose such a design method in this paper and begin its development in Section \ref{sec:proxy}.

    \section{A Proxy for the Joint Sampling Distribution}\label{sec:proxy} 

\subsection{Proxies to Posterior Probabilities}\label{sec:proxy.1}

We now create a proxy to the joint sampling distribution of $\boldsymbol{\tau}(\boldsymbol{x}_{n})$. These proxies motivate the theory that underpins the design methodology proposed in Section \ref{sec:power}. However, our methods simply adapt large-sample linear trends from these proxies to justify estimating the true sampling distribution of $\boldsymbol{\tau}(\boldsymbol{x}_{n})$ at only two values of $n$ using simulation. We create a proxy for the joint sampling distribution of posterior probabilities $\boldsymbol{\tau}_{\text{PP}}(\boldsymbol{x}_{n})$ in this subsection and augment this proxy to accommodate predictive probabilities in Section \ref{sec:proxy.2}. We emphasize adaptive designs with time-to-event outcomes, but the theory can be simplified for non-adaptive designs and other outcome types. 

For all settings we consider, our proxies rely on the asymptotic normality of the generalized posterior $\pi(\theta~|~\boldsymbol{x}_{n_t})$. This asymptotic normality should hold under the data-generation process $F_r \in \mathcal{F}$ for each simulation repetition $r$. We let $\hat{\theta}^{_{(n)}}_{t,r}$ be the M-estimate for the $t^{\text{th}}$ analysis. For large $n_t$, we require that the generalized posterior in (\ref{eq:M.post}) admits the \color{black} asymptotic \color{black} approximation
\begin{equation}\label{eq:M.BvM}
\pi(\theta~|~\boldsymbol{x}_{n_t, r}) \approx \mathcal{N}(\hat{\theta}^{_{(n)}}_{t,r}, \sigma^2_{t, r}/n_t),
\end{equation}
where $\sigma^2_{t, r}$ is the limiting \color{black} unscaled \color{black} variance for simulation repetition $r$. Our theoretical results rely on the existence of a variance $\sigma^2_{t, r}$ that does not depend on $n_t$, and we consider the expression for $\sigma^2_{t, r}$ in Section \ref{sec:ex}. The subscript $t$ appears in  $\sigma^2_{t, r}$ because we allow the treatment allocation probabilities to differ across the stages of the experiment, but these allocations must be pre-specified before collecting data. As we discuss in Section \ref{sec:disc}, our methodology thus precludes designs with response-adaptive randomization. 

The result in (\ref{eq:M.BvM}) holds true in many settings when the learning rate $\omega$ in (\ref{eq:M.post}) is a fixed constant. Regardless of the loss function that defines (\ref{eq:M.post}), the prior $\pi(\theta)$ must be continuous and positive at $\theta^*_r = \arg\min_{\theta}\mathbb{E}_{F_r}[\rho(\boldsymbol{X}, \theta)]$. For a broad class of loss functions such that $\rho(\boldsymbol{X}_i, \theta)$ is independent of $\rho(\boldsymbol{X}_j, \theta)$ for all observations $i \ne j$, \citet{marusic2025theoretical} proved the result in (\ref{eq:M.BvM}) under weak generalized local asymptotic normality assumptions; see Appendix B.1 of the online supplement for further discussion. 

Alternative theory may be required to apply our methodology in specialized settings. For example, the loss function in (\ref{eq:partial}) based on the partial likelihood is such that $\rho(\boldsymbol{X}_i, \theta)$ and $\rho(\boldsymbol{X}_j, \theta)$ are dependent if subject $j$ is in the risk set ${K}_i$. The result in (\ref{eq:M.BvM}) holds true for this loss function by applying the martingale central limit theorem along with a second-order Taylor series expansion (see e.g., \citet{fleming2005counting}). 

The following conditions must also hold true to invoke the result in (\ref{eq:M.BvM}) for adaptive designs with temporary censoring. First, subjects must enter into the experiment at a fixed rate such that we expect a given enrollment for each unit of time. Second, the $t^{\text{th}}$ analysis occurs after $n_t$ subjects have entered into the study. Third, there must be a fixed observation period (i.e., maximum follow-up time). These design assumptions are reasonable for many clinical trials. In Appendix B.1, we show the resulting expected proportion of outcomes that are temporarily censored at analysis $t$ is proportional to $1/n_t$ for large $n_t$; \color{black} we demonstrate how to assess the suitability of this asymptotic result for finite sample sizes using a quick simulation in Appendices E.1 and F.1. \color{black} 

In group sequential designs, the generalized posteriors of $\theta$ at distinct analyses are not independent because data from earlier stages are retained in subsequent ones. Our proxies account for this dependence via the joint sampling distribution of the M-estimator $\hat{\boldsymbol{\theta}}^{_{(n)}}_r = \{\hat{\theta}^{_{(n)}}_{t,r}\}_{t=1}^T$ across all analyses. We index all components of $\hat{\boldsymbol{\theta}}^{_{(n)}}_r$ by the sample size $n = n_1$ at the first analysis for simplicity, but note that $\hat{\theta}^{_{(n)}}_{t,r}$ is based on $n_t = c_tn$ observations. Our proxies also require that the approximate joint sampling distribution of the M-estimator, $\hat{\boldsymbol{\theta}}^{_{(n)}} ~|~ F = F_{r}$, for a hypothetical design without early stopping is
  \begin{equation}\label{eq:joint.M}
  \hat{\boldsymbol{\theta}}^{_{(n)}}_r \sim \mathcal{N}\left(\theta^*_r \times \mathbf{1}_T, \dfrac{1}{n}\times {\bf{V}}_r\right),
\end{equation}
where the matrix ${\bf{V}}_r$ does not depend on $n$.  

The $(i,j)$-entry of ${\bf{V}}_r$ is $v_{r,i,j} =$  $n\text{Cov}(\hat{\theta}^{_{(n)}}_{i,r}, \hat{\theta}^{_{(n)}}_{j, r})$ for $ 1 \le i, j \le T$. When the treatment allocations are pre-specified as discussed earlier, $v_{r,i,j}$ does not depend on $n$ since $\text{Corr}(\hat{\theta}^{_{(n)}}_{i,r}, \hat{\theta}^{_{(n)}}_{j, r})$, $\sqrt{n}\text{Var}(\hat{\theta}^{_{(n)}}_{i,r})^{1/2}$, and $\sqrt{n}\text{Var}(\hat{\theta}^{_{(n)}}_{j,r})^{1/2}$ are asymptotically independent of the sample size. 
As explained in Section \ref{sec:ex.1}, we cannot utilize the result in (\ref{eq:joint.M}) in the presence of temporary censoring \emph{and} misspecification of the estimating equation, $\sum_{i=1}^{n_t}\nabla_{\theta}\rho(\boldsymbol{X}_i, \theta) = 0$,  relative to $F_r$. \color{black} As a consequence of Appendices B.1 and B.2, $\hat{\boldsymbol{\theta}}^{_{(n)}}_r$ based on the data with temporary censoring differs from the estimator based on the fully followed data by $O_p(1/n)$. We therefore do not adjust the denominator of the variances in  (\ref{eq:M.BvM}) and (\ref{eq:joint.M}) to account for temporary censoring; the unadjusted variances are suitable asymptotically by Slutsky's theorem. \color{black}

 When  $\rho(\boldsymbol{X}_i, \theta)$ is independent of $\rho(\boldsymbol{X}_j, \theta)$ for all observations $i \ne j$, the result in (\ref{eq:joint.M}) holds true under the standard regularity conditions for M-estimators in \citet{huber2009robust}. Appendix B.1 contains further discussion. As for the result in (\ref{eq:M.BvM}), certain loss functions may require specialized theory, such as that \color{black} associated with sequential analysis based on the partial likelihood  \citep{tsiatis1981asymptotic, sellke1983sequential}.  \color{black} If practitioners cannot find asymptotic theory that readily applies to their loss function, they may confirm whether the approximations in (\ref{eq:M.BvM}) and (\ref{eq:joint.M}) hold true using a small-scale simulation that considers various large values of $n_t$. 


To develop our proxies, we use conditional cumulative distribution function (CDF) inversion to map realizations from the $T$-dimensional distribution in (\ref{eq:joint.M}) to points $\boldsymbol{u} = \{u_t \}_{t=1}^T \in [0,1]^T$. For repetition $r$, we obtain the first component $\hat{\theta}^{_{(n)}}_{1, r}$ of the M-estimate as the $u_{1}$-quantile of the sampling distribution of $\hat{\theta}^{_{(n)}}_{1} ~|~ F_{r}$. For the remaining components, we obtain $\hat{\theta}^{_{(n)}}_{t, r}$ as the $u_{t}$-quantile of the sampling distribution of $\hat{\theta}^{_{(n)}}_{t} ~|~ \{\hat{\theta}^{_{(n)}}_{s} = \hat{\theta}^{_{(n)}}_{s, r}\}_{s=1}^{t-1}, \hspace*{0.1pt} F_{r}$. Implementing this process with $R$ points $\{\boldsymbol{u}_{r}\}_{r = 1}^R \sim \mathcal{U}\left([0,1]^{T}\right)$ and data-generation processes $\{F_{r}\}_{r = 1}^R \sim \mathcal{F}$ yields a sample from the approximate sampling distribution of $\hat{\boldsymbol{\theta}}^{_{(n)}}$ according to $\mathcal{F}$. For theoretical purposes, we substitute this sample $\{ \hat{\boldsymbol{\theta}}^{_{(n)}}_{ r}\}_{r=1}^R$ into  (\ref{eq:M.BvM}) to yield a proxy sample of posterior probabilities. We approximate the probability in the $t^{\text{th}}$ row of $\boldsymbol{\tau}_{\text{PP}}(\boldsymbol{x}_{n})$ as
      \begin{equation}\label{eq:proxy}
\tau^{_{(n)}}_{\text{PP}, t, r} = 
   \Phi\left(\dfrac{\delta_U - \hat{\theta}^{_{(n)}}_{t, r}}{\sqrt{\sigma_{t,r}^2/c_tn}}\right) - \Phi\left(\dfrac{\delta_L - \hat{\theta}^{_{(n)}}_{t, r}}{\sqrt{\sigma_{t,r}^2/c_tn}}\right)
\end{equation} 
where $\Phi(\cdot)$ is the standard normal CDF. The collection of $\{\tau^{_{(n)}}_{\text{PP}, t, r}\}_{t = 1}^T$ values corresponding to $\{\boldsymbol{u}_{r}\}_{r = 1}^R \sim \mathcal{U}\left([0,1]^{T}\right)$ and $\{F_{r}\}_{r = 1}^R \sim \mathcal{F}$ defines our proxy to the joint sampling distribution of $\boldsymbol{\tau}_{\text{PP}}(\boldsymbol{x}_{n})$. 

Under the predictive approach, there are two sources of randomness in the proxy sampling distribution of $\boldsymbol{\tau}_{\text{PP}}(\boldsymbol{x}_{n})$. The first source is associated with $F_{r}$. The second source is related to the point $\boldsymbol{u}_{r}$ used to generate the M-estimate $\hat{\boldsymbol{\theta}}^{_{(n)}}_{ r}~|~F_{r}$, which serves as a conduit for the data $\boldsymbol{x}_{n, r}$. When conditioning on $\boldsymbol{u}_{r}$ and $F_{r}$, the value of $\tau^{_{(n)}}_{\text{PP}, t, r}~|~\color{black}\{\tau^{_{(n)}}_{\text{PP},s,r}\}_{s=1}^{t-1}$ \color{black} is not a stochastic quantity. For $t = 1$, the conditioning set is empty. \color{black} To simplify notation, we henceforth denote the conditioning set as $\tau^{_{(n)}}_{\text{PP},t-1,r}$ instead of $\{\tau^{_{(n)}}_{\text{PP},s,r}\}_{s=1}^{t-1}$. \color{black} Given values of $\boldsymbol{u}_{r}$ and $F_{r}$, $\tau^{_{(n)}}_{\text{PP}, t, r}~|~\tau^{_{(n)}}_{\text{PP}, t-1,r}$ based on (\ref{eq:proxy}) is thus a deterministic function of $n$. Lemma \ref{lem1} provides a structure for these deterministic functions when the results in (\ref{eq:M.BvM}) and (\ref{eq:joint.M}) hold true.

    \begin{lemma}\label{lem1}
    For any given $F_r \sim \mathcal{F}$, let the results in (\ref{eq:M.BvM}) and (\ref{eq:joint.M}) hold true. 
    We consider a given point $\boldsymbol{u}_{r} \in [0,1]^{T}$. Across design stages, we suppose allocation ratios are pre-specified and sample sizes are such that $\{n_t\}_{t=1}^T = n \times \{c_t\}_{t=1}^T$ for constants $c_1 = 1$ and $\{c_t\}_{t=2}^T > 1$. Under temporary censoring, we suppose the expected temporary censoring proportion at analysis $t$ is proportional to $c_t^{-1}/n$ and the estimating equation $\sum_{i=1}^{n_t}\nabla_{\theta}\rho(\boldsymbol{X}_i, \theta) = 0$ holds. For $t = 1, \dots, T$, the functions in (\ref{eq:proxy}) are such that 
    \begin{equation}\label{eq:lem1}
        \tau^{_{(n)}}_{\emph{PP}, t, r}~|~\tau^{_{(n)}}_{\emph{PP}, t-1,r} = \Phi\left(
           f_{\emph{PP}, t}(\delta_U, \theta^*_{r})\sqrt{n} + g_{\emph{PP}, t}(\boldsymbol{u}_{r}) 
         \right) - \Phi\left(
           f_{\emph{PP}, t}(\delta_L, \theta^*_{r})\sqrt{n} + g_{\emph{PP}, t}(\boldsymbol{u}_{r}) 
         \right),
    \end{equation} where $f_{\emph{PP}, t}(\cdot)$ and  $g_{\emph{PP}, t}(\cdot)$ are functions that do not depend on $n$.
\end{lemma} 

We prove Lemma \ref{lem1} and derive expressions for $\tau^{_{(n)}}_{\text{PP}, t, r}~|~\tau^{_{(n)}}_{\text{PP}, t-1,r}$ in Appendix B.1. All lemmas in this section are used to prove new theoretical results about our proxy sampling distributions in Section \ref{sec:proxy.3} that allow us to greatly expedite the design of experiments based on generalized posteriors. 

\subsection{Proxies to Interim and Final Predictive Probabilities}\label{sec:proxy.2}

We next state Lemmas \ref{lem2} and \ref{lem3}, which pertain to proxies for the sampling distributions of $\boldsymbol{\tau}_{\text{IP}}(\boldsymbol{x}_{n})$ and $\boldsymbol{\tau}_{\text{FP}}(\boldsymbol{x}_{n})$. These proxies also are based on the points $\{\boldsymbol{u}_{r}\}_{r = 1}^R$ and data-generation processes $\{F_{r}\}_{r = 1}^R$ from Section \ref{sec:proxy.1}, so we construct a proxy to the \emph{joint} sampling distribution of $\boldsymbol{\tau}_{\text{PP}}(\boldsymbol{x}_{n})$, $\boldsymbol{\tau}_{\text{IP}}(\boldsymbol{x}_{n})$, and $\boldsymbol{\tau}_{\text{FP}}(\boldsymbol{x}_{n})$. Because the structure of the derivations for the proxies to the interim and final predictive probabilities is similar to that for $\boldsymbol{\tau}_{\text{PP}}(\boldsymbol{x}_{n})$, full derivations are provided in Appendices \color{black}B.2 and B.3 \color{black} of the supplement. Here, we overview the required conditions for these lemmas. 

We introduce Lemmas \ref{lem2} and \ref{lem3} for an adaptive design with temporary censoring at analysis $t < T$.  However, Lemma \ref{lem3} for $\boldsymbol{\tau}_{\text{FP}}(\boldsymbol{x}_{n})$ can be simplified for designs without temporary censoring as described in Appendix B.3. Both Lemmas \ref{lem2} and \ref{lem3} hold true under the assumptions for Lemma \ref{lem1} and \color{black}two additional conditions. First, \color{black}the posterior predictive distribution $\dot{F}$ in (\ref{eq:ppd}) must be defined via $\tilde{F}_r$ such that $\mathbb{E}_{\tilde{F}_r|\hat{\theta}^{_{(n)}}_{t,r}}[\nabla_{\theta}\rho(\tilde{X}, \hat{\theta}^{_{(n)}}_{t,r})~|~\boldsymbol{X}_n] = \color{black}O_p(n^{-1})\color{black}$, where $\tilde{X}$ is an observation from $\dot{F}$ such that $\tilde{X}_i = X_i$ if outcome $i$ is observed at analysis $t$. This condition ensures that the data-generation process $\tilde{F}_r(\tilde{x}~|~\hat{\theta}^{_{(n)}}_{t,r}, \boldsymbol{x}_{n_t, r})$ does not introduce asymptotic bias into the estimation process for $\theta$; we discuss why this assumption is reasonable when using generalized posteriors in Sections \ref{sec:ex} and \ref{sec:orvac}. \color{black}Second, the expected number of temporarily censored outcomes must be sufficiently large as detailed in Appendix B.2.

\color{black}
In Appendix B.2, we also derive proxies \color{black} $\tau^{_{(n)}}_{\text{IP}, t, r}~|~\{\tau^{_{(n)}}_{\text{PP}, s, r}\}_{s=1}^t$ \color{black} that model conditional dependence in the proxy sampling distribution since $\tau_{\text{IP}, t}(\boldsymbol{x}_n)$ is defined by conditioning on the observed data $\boldsymbol{x}_{n_t}$. \color{black}We henceforth use simplified notation for the conditioning set: \color{black} $\tau^{_{(n)}}_{\text{IP}, t, r}~|~\tau^{_{(n)}}_{\text{PP}, t, r}$\color{black}. Using logic from Section \ref{sec:proxy.1}, $\tau^{_{(n)}}_{\text{IP}, t, r}~|~\tau^{_{(n)}}_{\text{PP}, t,r}$ is a deterministic function of $n$ given $\boldsymbol{u}_{r}$ and $F_{r}$. Lemma \ref{lem2} provides a standard form for these functions. 

        \begin{lemma}\label{lem2}
    We suppose the conditions for Lemma \ref{lem1} are satisfied \color{black}and the expected number of temporarily censored outcomes is sufficiently large. \color{black} Let $\dot{F}$ in (\ref{eq:ppd}) be such that $\mathbb{E}_{\tilde{F}_r|\hat{\theta}^{_{(n)}}_{t,r}}[\nabla_{\theta}\rho(\tilde{X}, \hat{\theta}^{_{(n)}}_{t,r})~|~\boldsymbol{X}_n] = \color{black}O_p(n^{-1})\color{black}$. We consider a given point $\boldsymbol{u}_{r} \in [0,1]^{T}$ and data-generation process $F_{r}$. For $t = 1, \dots, T-1$, the proxy  functions in (B.15) of Appendix B.2 are such that 
     \begin{equation}\label{eq:lem2}
     \begin{split}
     \tau^{_{(n)}}_{\emph{IP}, t, r}~|~\tau^{_{(n)}}_{\emph{PP}, t,r} = \Phi\left(
           f_{\emph{IP}, t}(\delta_U, \theta^*_{r})n + g_{\emph{IP}, t}(\boldsymbol{u}_{r})\sqrt{n} 
         \right) -  \Phi\left(
           f_{\emph{IP}, t}(\delta_L, \theta^*_{r})n + \color{black}j_{\emph{IP}, t}\color{black}(\boldsymbol{u}_{r})\sqrt{n} 
         \right),
    \end{split}
    \end{equation}
    where $f_{\emph{IP}, t}(\cdot)$, $g_{\emph{IP}, t}(\cdot)$, \color{black} and $j_{\emph{IP}, t}(\cdot)$ \color{black} are functions that do not depend on $n$.
\end{lemma}

We prove this lemma in Appendix B.2 and now compare it to Lemma \ref{lem1}. The function  $f_{\text{IP}, t}(\cdot)$ is multiplied by $n$, whereas $f_{\text{PP}, t}(\cdot)$ in (\ref{eq:lem1}) was multiplied by $\sqrt{n}$. At a high level, this discrepancy occurs because the interim predictive probabilities condition on $\boldsymbol{x}_{n_t}$. Under the conditions in Lemma \ref{lem1}, the \color{black}proportion \color{black} of temporarily censored outcomes is proportional to $1/n_t$ for large $n_t$. The variation remaining in the temporarily censored outcomes thus decreases faster than the variation in $\pi(\theta~|~\tilde{\boldsymbol{x}}_{n_t})$. Similar justification applies to why $g_{\text{IP}, t}(\cdot)$  \color{black} and $j_{\text{IP}, t}(\cdot)$ \color{black}  -- but not $g_{\text{PP}, t}(\cdot)$ in (\ref{eq:lem1}) -- are multiplied by $\sqrt{n}$. Formal mathematical justification is provided in Appendix B.2.

In Appendix B.3, we derive proxies \color{black} $\tau^{_{(n)}}_{\text{FP}, t, r}~|~\{\tau^{_{(n)}}_{\text{PP}, s, r}\}_{s=1}^t$ \color{black} that model conditional dependence in the proxy sampling distribution. \color{black}The conditional proxies are $\tau^{_{(n)}}_{\text{FP}, t, r}~|~\tau^{_{(n)}}_{\text{PP}, t, r}$ in our simplified notation. \color{black} When conditioning on $\boldsymbol{u}_{r}$ and $F_{r}$, $\tau^{_{(n)}}_{\text{FP}, t, r}~|~\tau^{_{(n)}}_{\text{PP}, t,r}$ is also a deterministic function of $n$. Lemma \ref{lem3}, proven in Appendix B.3, provides a standard form for these functions. 

        \begin{lemma}\label{lem3}
    We suppose the conditions for Lemmas \ref{lem1} and \ref{lem2} are satisfied. We consider a given point $\boldsymbol{u}_{r} \in [0,1]^{T}$ and data-generation process $F_{r}$. For $t = 1, \dots, T-1$, the proxy functions in (B.19) of Appendix B.3 are such that 
     \begin{equation}\label{eq:lem3}
     \begin{split}
     \tau^{_{(n)}}_{\emph{FP}, t, r}~|~\tau^{_{(n)}}_{\emph{PP}, t,r} &= \Phi\left(\color{black}s_t(n)\color{black}\{
           f_{\emph{FP}, t}(\delta_U, \theta^*_{r})\sqrt{n} + g_{\emph{FP}, t}(\boldsymbol{u}_{r})\} 
         \right) - \\
         & ~~~~~\Phi\left(
           \color{black}s_t(n)\color{black}\{f_{\emph{FP}, t}(\delta_L, \theta^*_{r})\sqrt{n} + \color{black}j_{\emph{FP}, t}\color{black}(\boldsymbol{u}_{r})\} 
         \right),
    \end{split}
    \end{equation}
    where $f_{\emph{FP}, t}(\cdot)$, $g_{\emph{FP}, t}(\cdot)$, \color{black}and $j_{\emph{FP}, t}(\cdot)$ are functions that do not depend on $n$, and $s_t(n) = 1 + O(n^{-1/2})$. \color{black}
\end{lemma}

We now briefly comment on Lemma \ref{lem3}. Similar to $f_{\text{PP}, t}(\cdot)$ in (\ref{eq:lem1}), $f_{\text{FP}, t}(\cdot)$ is multiplied by $\sqrt{n}$. While the final predictive probabilities also condition on $\boldsymbol{x}_{n_t}$, the variability in the outcomes from future stages is proportional to the variability in $\pi(\theta~|~\tilde{\boldsymbol{x}}_{n_T})$ under the conditions in Lemma \ref{lem1}. \color{black} The function $s_t(n)$ reflects the variability in the outcomes that are temporarily censored (if applicable). In Appendix B.3, we show that $s_t(n)$ converges to 1 as $n \rightarrow \infty$ \color{black} because the variability in the temporarily censored outcomes decreases faster with $n$ than the variability in the future-stage outcomes.

\subsection{Implications for Computationally Efficient Design}\label{sec:proxy.3}

    Our proxy to the sampling distribution of $\boldsymbol{\tau}(\boldsymbol{x}_{n})$ relies on asymptotic results, so it may differ from the true sampling distribution for finite $n$. As such, this proxy only motivates our theoretical result in Theorem \ref{thm1}, which utilizes the deterministic functions derived in Lemmas \ref{lem1}, \ref{lem2}, and \ref{lem3}. Theorem \ref{thm1} guarantees that the logits of $\tau^{_{(n)}}_{\text{PP}, t, r}~|~\tau^{_{(n)}}_{\text{PP}, t-1,r}$, $\tau^{_{(n)}}_{\text{IP}, t, r}~|~\tau^{_{(n)}}_{\text{PP}, t,r}$, and  $\tau^{_{(n)}}_{\text{FP}, t, r}~|~\tau^{_{(n)}}_{\text{PP}, t,r}$ are approximately linear functions of the sample size (i.e., linear functions of $n$ or $n^2$). We later adapt this result to estimate the operating characteristics of adaptive designs across a wide range of sample sizes by estimating the true sampling distribution of $\boldsymbol{\tau}(\boldsymbol{x}_{n})$ at only two values of $n$.

    \begin{theorem}\label{thm1}
     We suppose the conditions for Lemmas \ref{lem1}, \ref{lem2}, and \ref{lem3} are satisfied. Define $\emph{logit}(x) = \emph{log}(x) - \emph{log}(1-x)$. We consider a given point $\boldsymbol{u}_{r} \in [0,1]^{T}$ and data-generation process $F_{r}$. Let $\Delta_r = (0.5 - \mathbb{I}\{\theta^*_r \notin (\delta_{L}, \delta_{U})\})$. The functions  $\{\tau^{_{(n)}}_{\emph{PP}, t, r}~|~\tau^{_{(n)}}_{\emph{PP}, t-1,r}\}_{t=1}^T$ based on (\ref{eq:lem1}), $\{\tau^{_{(n)}}_{\emph{IP}, t, r}~|~\tau^{_{(n)}}_{\emph{PP}, t,r}\}_{t=1}^{T-1}$ based on (\ref{eq:lem2}), and $\{\tau^{_{(n)}}_{\emph{FP}, t, r}~|~\tau^{_{(n)}}_{\emph{PP}, t,r}\}_{t=1}^{T-1}$ based on (\ref{eq:lem3}) are such that
 \begin{enumerate}
     \item[(a)] $\lim\limits_{n \rightarrow \infty} \dfrac{d}{dn}~\emph{logit}\left(\tau^{_{(n)}}_{\emph{PP}, t, r}~|~\tau^{_{(n)}}_{\emph{PP}, t-1,r}\right)= \Delta_r\times\emph{min}\{f_{\emph{PP}, t}(\delta_{U}, \theta^*_r)^2, f_{\emph{PP}, t}(\delta_{L}, \theta^*_r)^2\} $. 
     \item[(b)] $\lim\limits_{n \rightarrow \infty} \dfrac{d}{d(n^2)}~\emph{logit}\left(\tau^{_{(n)}}_{\emph{IP}, t, r}~|~\tau^{_{(n)}}_{\emph{PP}, t,r}\right)= \Delta_r\times\emph{min}\{f_{\emph{IP},t}(\delta_{U}, \theta^*_r)^2, f_{\emph{IP},t}(\delta_{L}, \theta^*_r)^2\} $. 
     \item[(c)] $\lim\limits_{n \rightarrow \infty} \dfrac{d}{dn}~\emph{logit}\left(\tau^{_{(n)}}_{\emph{FP}, t, r}~|~\tau^{_{(n)}}_{\emph{PP}, t,r}\right)=  \Delta_r\times\emph{min}\{f_{\emph{FP},t}(\delta_{U}, \theta^*_r)^2, f_{\emph{FP},t}(\delta_{L}, \theta^*_r)^2\} $. 
 \end{enumerate}
\end{theorem} 

We prove Theorem \ref{thm1} in Appendix C of the supplement. 
The practical implications of Theorem \ref{thm1} are as follows. The limiting derivatives in parts $(a)$, $(b)$, and $(c)$ are constants that do not depend on $n$. These limiting derivatives also do not depend on the point $\boldsymbol{u}_{r}$ which controls the dependence within the joint proxy sampling distribution of $\boldsymbol{\tau}(\boldsymbol{x}_{n})$. The limiting derivatives of (i) $\text{logit}(\tau^{_{(n)}}_{\text{PP}, t, r})$ and $\text{logit}(\tau^{_{(n)}}_{\text{PP}, t, r}~|~\tau^{_{(n)}}_{\text{PP}, t-1,r})$, (ii) $\text{logit}(\tau^{_{(n)}}_{\text{IP}, t, r})$ and $\text{logit}(\tau^{_{(n)}}_{\text{IP}, t, r}~|~\tau^{_{(n)}}_{\text{PP}, t,r})$, and (iii) $\text{logit}(\tau^{_{(n)}}_{\text{FP}, t, r})$ and $\text{logit}(\tau^{_{(n)}}_{\text{FP}, t, r}~|~\tau^{_{(n)}}_{\text{PP}, t,r})$ are therefore the same. In the joint proxy sampling distribution, the linear approximations to $l^{_{(n)}}_{\text{PP}, t, r} = \text{logit}(\tau^{_{(n)}}_{\text{PP}, t, r}~|~\tau^{_{(n)}}_{\text{PP}, t-1,r})$ and $l^{_{(n)}}_{\text{FP},t, r} = \text{logit}(\tau^{_{(n)}}_{\text{FP},t, r}~|~\tau^{_{(n)}}_{\text{PP}, t,r})$ as functions of $n$ are thus good global approximations for large sample sizes; we draw a similar conclusion about the linear approximation to $l^{_{(n)}}_{\text{IP}, t, r} = \text{logit}(\tau^{_{(n)}}_{\text{IP}, t, r}~|~\tau^{_{(n)}}_{\text{PP}, t,r})$ as a function of $n^2$. These linear approximations should be locally suitable for a range of smaller sample sizes. 

\color{black}
Under the conditional approach where $\{F_{r}\}_{r=1}^R$ are the same, the (conditional) quantiles of the sampling distributions of $l^{_{(n)}}_{\text{PP}, t, r}$, $l^{_{(n)}}_{\text{IP},t, r}$, and $l^{_{(n)}}_{\text{FP},t, r}$ therefore change linearly as a function of $n$ or $n^2$. Given $F_r$, the slopes of these linear functions in the proxy distribution are the same for \emph{all} quantiles. This insight even applies in cases where both $\delta_L$ and $\delta_U$ are finite since the limiting slope is dictated by the interval endpoint that is closer to $\theta^*_r$. \color{black}In Section \ref{sec:power}, we exploit and adapt these linear trends in the proxy sampling distribution to flexibly model the logits of posterior summaries as functions of $n$ when estimating true sampling distributions of $\boldsymbol{\tau}(\boldsymbol{x}_{n})$ under the conditional or predictive approach. 

   \section{A Computationally Efficient Design Procedure}\label{sec:power}

   \subsection{Stopping Rules in Adaptive Designs}\label{sec:power.1}

      We now introduce example stopping rules based on $\boldsymbol{\tau}(\boldsymbol{x}_{n})$ that inform interim decisions in group sequential designs with early stopping. As mentioned in Section \ref{sec:methods.2}, an ongoing experiment could be stopped for success at analysis $t$ if $\tau_{\text{PP},t}(\boldsymbol{x}_{n}) \ge \gamma_t$, where $\boldsymbol{\gamma}=\{\gamma_t\}_{t=1}^T \in [0,1]^T$ are success thresholds. For designs with time-to-event outcomes and temporary censoring, it may instead be sensible to make decisions based on interim predictive probabilities. For example, an ongoing experiment could be stopped for success at the $t^{th}$ analysis if $\tau_{\text{IP},t}(\boldsymbol{x}_{n}) \ge \xi_t$ such that $\boldsymbol{\xi}=\{\xi_t\}_{t=1}^{T-1} \in [0,1]^{T-1}$ are a distinct set of success thresholds. For designs with or without temporary censoring, final predictive probabilities may be used to stop an ongoing experiment for failure if $\tau_{\text{FP},t}(\boldsymbol{x}_{n}) < \kappa_t$, where  $\boldsymbol{\kappa}=\{\kappa_t\}_{t=1}^{T-1} \in [0,1]^{T-1}$ are failure thresholds. 

   To assess operating characteristics of designs with general stopping rules, we define a binary indicator $\nu(\boldsymbol{x}_{n})$ that equals 1 if and only if the experiment stops for \emph{success} at any analysis $t = 1, \dots, T$. An example design with $T=2$ analyses and temporary censoring emphasizes the relationship between $\boldsymbol{\tau}(\boldsymbol{x}_{n})$ and $\nu(\boldsymbol{x}_{n})$. At the first analysis, this example design considers stopping for success based on $\boldsymbol{\tau}_{\text{IP}}(\boldsymbol{x}_{n})$ before considering stopping for failure based on $\boldsymbol{\tau}_{\text{FP}}(\boldsymbol{x}_{n})$. Only stopping for success based on $\boldsymbol{\tau}_{\text{PP}}(\boldsymbol{x}_{n})$ is considered at the second analysis. We have that $\nu(\boldsymbol{x}_{n}) = 1$ based on the first analysis if and only if $\tau_{\text{IP}, 1}(\boldsymbol{x}_{n}) \ge \xi_1$. Moreover, $\nu(\boldsymbol{x}_{n}) = 1$ based on analysis 2 if and only if $\tau_{\text{IP}, 1}(\boldsymbol{x}_{n}) < \xi_1$, $\tau_{\text{FP},1}(\boldsymbol{x}_{n}) \ge \kappa_1$, and $\tau_{\text{PP}, 2}(\boldsymbol{x}_{n}) \ge \gamma_2$.

We define operating characteristics with respect to the data-generation family $\mathcal{F}$. For a given family $\mathcal{F}$, the probability of stopping for success across all analyses is $\mathbb{E}_{\mathcal{F}}[\Pr(\nu(\boldsymbol{x}_{n}) = 1)~|~F]$. Given the simulation results introduced in Section \ref{sec:methods.2}, this probability is estimated as 
  \begin{equation}\label{eq:power}
  \dfrac{1}{R}\sum_{r=1}^R \mathbb{I}\left\{\nu(\boldsymbol{x}_{n, r}) = 1\right\},
\end{equation} 
where $\boldsymbol{x}_{n, r}$ are generated using the data-generation process $F_r$ from $\mathcal{F}$. The power to correctly stop for success is \textbf{$\mathbb{E}_{\mathcal{F}_1}[\Pr(\nu(\boldsymbol{x}_{n}) = 1)~|~F]$}, where $\mathcal{F}_1$ is a data-generation family such that $H_1$ is true. We estimate power using (\ref{eq:power}) when $\{\boldsymbol{x}_{n_t, r}\}_{t = 1}^T$ are generated using $F_r$ obtained via $\mathcal{F}_1$. The frequentist type I error rate related to incorrectly stopping for success is $\mathbb{E}_{\mathcal{F}_0}[\Pr(\nu(\boldsymbol{x}_{n}) = 1)~|~F]$ where $\mathcal{F}_0$ is a data-generation family such that $H_0$ is true. Using $\mathcal{F}_0$, this probability can be estimated as in (\ref{eq:power}). 

Under the following conditions, power and the type I error rate for adaptive designs can be computed using an estimate of the sampling distribution $\{\boldsymbol{\tau}(\boldsymbol{x}_{n, r})\}_{r=1}^R$ for a hypothetical design without early stopping. First, the proportion of observations in each stage of the design must be pre-specified (as enforced by the constraint $\{n_t\}_{t=1}^T = n \times \{c_t\}_{t=1}^T$ for constants $c_1=1$ and $\{c_t\}_{t=2}^T > 1$). Second, the allocation ratios in each stage must be fixed before collecting data (i.e., response-adaptive randomization is not permitted). If the experiment stops early in a given simulation repetition $r$, we can simply ignore the components of $\boldsymbol{\tau}(\boldsymbol{x}_{n, r})$ that correspond to subsequent analyses when computing operating characteristics.

   \subsection{Efficient Design Under the Hybrid Approach}\label{sec:power.2}

   In Algorithm \ref{alg1}, we generalize the results from Theorem \ref{thm1} to develop a hybrid design procedure for experiments based on generalized posteriors. This procedure chooses sample sizes and tunes decision thresholds to ensure criteria for the operating characteristics are satisfied; it requires that we estimate the sampling distribution of $\boldsymbol{\tau}(\boldsymbol{x}_{n})$ by simulating data $\boldsymbol{x}_n$ at \emph{only} two values of $n$: $n_{a}$ and $n_b$. The initial sample size for the first analysis $n_a$ in an adaptive design can be selected based on the anticipated budget for the experiment. In Algorithm \ref{alg1}, we add a subscript to $\boldsymbol{x}_{n,r}$ between $n$ and $r$ that distinguishes whether the data are generated according to $\mathcal{F}_0$ or $\mathcal{F}_1$ defined in Section \ref{sec:power.1}. We also define criteria for the error probabilities. Under $\mathcal{F}_1$ where $H_1$ is true, we want $\mathbb{E}_{\mathcal{F}_1}[\Pr(\nu(\boldsymbol{x}_{n}) = 1)~|~F] \ge 1 - \beta$. We want $\mathbb{E}_{\mathcal{F}_0}[\Pr(\nu(\boldsymbol{x}_{n}) = 1)~|~F]  \le \alpha$ under $\mathcal{F}_0$ where $H_0$ is true. Algorithm \ref{alg1} details a general application of our methodology for an adaptive design with the conditional approach. We later describe potential modifications, and we provide a simpler algorithm for non-adaptive designs in Appendix D.1 of the supplement.

         \begin{algorithm}
\caption{Procedure to Determine Sample Sizes and Decision Thresholds}
\label{alg1}

\begin{algorithmic}[1]
\Procedure{Design}{$\rho$, $\pi(\theta)$, $W$, $\delta_L$, $\delta_U$, $\mathcal{F}_0$, $\mathcal{F}_1$, $R$, $M$, $n_a$, $\{c_t\}_{t=1}^T$, $\alpha$, $\beta$}
\State  Compute $\{\boldsymbol{\tau}(\boldsymbol{x}_{n_a, 0, r})\}_{r = 1}^R$ obtained with $F_r \sim \mathcal{F}_0$. 
\State Choose thresholds from $\boldsymbol{\gamma}$, $\boldsymbol{\xi}$, and $\boldsymbol{\kappa}$ to ensure $R^{-1}\sum_{r=1}^R\mathbb{I}\{\nu(\boldsymbol{x}_{n_a, 0, r}) = 1\} \le \alpha$.
\State  Compute $\{\boldsymbol{\tau}(\boldsymbol{x}_{n_a, 1, r})\}_{r = 1}^R$ obtained with $F_r \sim \mathcal{F}_1$. 
\State If $R^{-1}\sum_{r=1}^R\mathbb{I}\{\nu(\boldsymbol{x}_{n_a, 1, r}) = 1\} \ge 1 - \beta$, choose $n_b < n_a$. If not, choose $n_b > n_a$.
\State  Compute $\{\boldsymbol{\tau}(\boldsymbol{x}_{n_b, 1, r})\}_{r = 1}^R$ obtained with $F_r \sim \mathcal{F}_1$.
\For{$d$ in $1$:$R$}
 \For{$t$ in $1$:$T$}
 \State Let $\boldsymbol{x}_{n_a, 1, r}$ correspond to the $d^{\text{th}}$ order statistic of $\{l_{\text{PP}, t}(\boldsymbol{x}_{n_a, 1, r})\}_{r = 1}^R$. 
\State Pair the $d^{\text{th}}$ order statistics of $\{l_{\text{PP}, t}(\boldsymbol{x}_{n_a, 1, r})\}_{r = 1}^R$ and $\{l_{\text{PP}, t}(\boldsymbol{x}_{n_b, 1, r})\}_{r = 1}^R$ with  linear  approxi-  \linebreak \hspace*{42pt} mations   to obtain $\hat{l}_{\text{PP}, t}(\boldsymbol{x}_{n, 1, r})$ estimates for new $n$.
\If {$t < T$}
 \State Modify Lines 9 and 10 with $\{l_{\text{IP}, t}(\boldsymbol{x}_{n_a, 1, r})\}_{r = 1}^R$ and $\{l_{\text{IP}, t}(\boldsymbol{x}_{n_b, 1, r})\}_{r = 1}^R$  to  estimate \linebreak \hspace*{57pt} $\hat{l}_{\text{IP}, t}(\boldsymbol{x}_{n, 1, r})$ as linear functions of $n^2$  for new $n$.
  \State Repeat Lines 9 and 10 with $\{l_{\text{FP}, t}(\boldsymbol{x}_{n_a, 1, r})\}_{r = 1}^R$ and $\{l_{\text{FP}, t}(\boldsymbol{x}_{n_b, 1, r})\}_{r = 1}^R$  to  estimate \linebreak \hspace*{57pt}  $\hat{l}_{\text{FP}, t}(\boldsymbol{x}_{n, 1, r})$ for new $n$.
 \EndIf
 \EndFor
 \EndFor
\State Obtain $\{\hat{\boldsymbol{\tau}}(\boldsymbol{x}_{n, 1, r})\}_{r = 1}^R$ as the expits of  $\{\hat{l}_{\text{PP}, t}(\boldsymbol{x}_{n, 1, r})\}_{t = 1}^{T}$, $\{\hat{l}_{\text{IP}, t}(\boldsymbol{x}_{n, 1, r})\}_{t = 1}^{T-1}$, and  $\{\hat{l}_{\text{FP}, t}(\boldsymbol{x}_{n, 1, r})\}_{t = 1}^{T-1}$.
\State Find $n_c$, the smallest $n \in \mathbb{Z}^+$ such that $R^{-1}\sum_{r=1}^R\mathbb{I}\{\hat{\nu}(\boldsymbol{x}_{n, 1, r}) = 1\} \ge 1 - \beta$.
 \State \Return $n_c$ as recommended $n$ and $\{\boldsymbol{\gamma}, \boldsymbol{\xi}, \boldsymbol{\kappa}\}$ as recommended decision thresholds

\EndProcedure

\end{algorithmic}
\end{algorithm}

We now elaborate on several steps of Algorithm \ref{alg1}. In Line 3, we choose suitable vectors for the relevant decision thresholds to ensure the estimate for $\mathbb{E}_{\mathcal{F}_0}[\Pr(\nu(\boldsymbol{x}_{n}) = 1)~|~F]$ based on (\ref{eq:power}) is at most $\alpha$. Initial values for the success thresholds $\boldsymbol{\gamma}$ can be chosen using standard theory from group sequential designs \citep{jennison1999group}; the remaining thresholds $\boldsymbol{\xi}$ and $\boldsymbol{\kappa}$ can be initialized as high and low probabilities, respectively. While implementing Algorithm \ref{alg1}, the choices for $\boldsymbol{\gamma}$, $\boldsymbol{\xi}$,  and $\boldsymbol{\kappa}$ may be tuned using sampling distribution estimates obtained under $\mathcal{F}_0$ and $\mathcal{F}_1$. Since Algorithm \ref{alg1} efficiently estimates sampling distributions across a broad range of sample sizes, the proposed methodology streamlines this tuning process. 

All posterior summaries approximated in Lines 2 to 6 of Algorithm \ref{alg1} are obtained by simulating data according to $\mathcal{F}_0$ or $\mathcal{F}_1$. We compute logits of these summaries under $\mathcal{F}_1$: $l_{\text{PP}, t}(\boldsymbol{x}_{n, 1, r}) = \text{logit}(\tau_{\text{PP}, t}(\boldsymbol{x}_{n, 1, r}))$, $l_{\text{IP}, t}(\boldsymbol{x}_{n, 1, r}) = \text{logit}(\tau_{\text{IP}, t}(\boldsymbol{x}_{n, 1, r}))$, and $l_{\text{FP}, t}(\boldsymbol{x}_{n, 1, r}) = \text{logit}(\tau_{\text{FP}, t}(\boldsymbol{x}_{n, 1, r}))$. If not all components of $\boldsymbol{\tau}(\boldsymbol{x}_{n})$ inform decision rules in a particular design, various rows in $\boldsymbol{\tau}(\boldsymbol{x}_{n})$ may be ignored. We recommend calculating posterior summaries using nonparametric kernel density estimates so that these logits are finite; \color{black} to ensure the logits of posterior predictive probabilities are finite, a density estimate of the logits of the $M$ $\Pr(H_1~|~\tilde{\boldsymbol{x}}_{n_t, m})$ values from Section \ref{sec:methods.2} can be compared to $\text{logit}(\gamma_t)$. \color{black} 

We construct linear approximations separately for each summary using logits corresponding to sample sizes $n_a$ and $n_b$ under the model $\mathcal{F}_1$ in Lines 7 to 13; we underscore that the approximations in Line 12 are linear functions of $n^2$. This second sample size $n_b$ could be chosen as the smallest sample size $n$ such that power is sufficiently large when the sampling distribution of $\boldsymbol{\tau}(\boldsymbol{x}_{n})$ is modeled using lines that pass through $(n_a, \hat{l}_{\text{PP}, t}(\boldsymbol{x}_{n_a, 1, r}))$, $(n_a^2, \hat{l}_{\text{IP}, t}(\boldsymbol{x}_{n_a, 1, r}))$, or $(n_a, \hat{l}_{\text{FP}, t}(\boldsymbol{x}_{n_a, 1, r}))$ with the limiting slopes from Theorem \ref{thm1} (see Appendix D.1 for details). Alternatively, $n_b$ could be chosen to explore a range of relevant sample sizes between $n_a$ and $n_b$. We use these linear approximations to estimate logits of posterior summaries for new values of $n$ as $\hat{l}_{\text{PP}, t}(\boldsymbol{x}_{n, 1, r})$, $\hat{l}_{\text{IP}, t}(\boldsymbol{x}_{n, 1, r})$, or $\hat{l}_{\text{FP}, t}(\boldsymbol{x}_{n, 1, r})$. 
To maintain the proper level of dependence in the joint sampling distribution of $\boldsymbol{\tau}(\boldsymbol{x}_{n})$, we group the linear functions from Lines 10 to 13 across all posterior summaries based on the sample $\boldsymbol{x}_{n_a, 1, r}$ that defined the linear approximations; please see Appendix D.2 for an explanatory figure.

Given the linear trends in the proxy sampling distribution quantiles discussed in Section \ref{sec:proxy.3}, these linear approximations can be constructed based on order statistics of estimates of the true sampling distributions under the conditional approach. Because Theorem \ref{thm1} ensures that the limiting slopes of (i) $\text{logit}(\tau^{_{(n)}}_{\text{PP}, t, r})$ and $\text{logit}(\tau^{_{(n)}}_{\text{PP},t, r}~|~\tau^{_{(n)}}_{\text{PP},t-1,r})$, (ii) $\text{logit}(\tau^{_{(n)}}_{\text{IP}, t, r})$ and $\text{logit}(\tau^{_{(n)}}_{\text{IP}, t, r}~|~\tau^{_{(n)}}_{\text{PP},t,r})$, and (iii) $\text{logit}(\tau^{_{(n)}}_{\text{FP}, t, r})$ and $\text{logit}(\tau^{_{(n)}}_{\text{FP}, t, r}~|~\tau^{_{(n)}}_{\text{PP},t,r})$ from the proxy sampling distribution are the same, we use the \emph{marginal} sampling distributions for each row of $\boldsymbol{\tau}(\boldsymbol{x}_n)$ to estimate slopes for the \emph{conditional} logits. 

Under the predictive approach, we modify the process in Lines 7 to 13. We split the logits of the posterior summaries for each $n$ value into subgroups based on the order statistics of their $\theta^*_{r}$ values before constructing the linear approximations. \color{black} This process ensures the slopes of the linear approximations are similar within each bin, allowing us to construct the approximations using order statistics. \color{black} In Line 15, we find the smallest value of $n$ such that our estimate for $\mathbb{E}_{\mathcal{F}_1}[\Pr(\nu(\boldsymbol{x}_{n}) = 1)~|~F]$ based on the indicators $\{\hat{\boldsymbol{\nu}}(\boldsymbol{x}_{n, 1, r})\}_{r = 1}^R$ that correspond to $\{\hat{\boldsymbol{\tau}}(\boldsymbol{x}_{n, 1, r})\}_{r = 1}^R$ is at least $1 - \beta$. Sample sizes throughout the group sequential design are obtained using the constants $\{c_t\}_{t=2}^T$.

 We did not estimate the sampling distribution of $\boldsymbol{\tau}(\boldsymbol{x}_{n_b})$ under $\mathcal{F}_0$ in Algorithm \ref{alg1}. To consider the worst-case operating characteristics, it is common practice to consider families $\mathcal{F}_0$ that assign all weight to data-generation processes $F_r$ such that $\theta^*_r$ equals $\delta_{L}$ or $\delta_{U}$. We show that all limiting slopes in  Theorem \ref{thm1} are zero for such models $\mathcal{F}_0$ in Appendix C. \color{black} Although the limiting derivative with respect to $n^2$ in part $(b)$ of Theorem \ref{thm1} is 0 in this case, we show that the limit of the derivative of  $\text{logit}(\tau^{_{(n)}}_{\text{IP}, t, r}~|~\tau^{_{(n)}}_{\text{PP},t,r})$ with respect to $n$ is a non-zero constant when $\theta^*_r$ equals $\delta_L$ or $\delta_U$ in Appendix C.2. In designs without interim predictive probabilities, \color{black} the type I error rate is thus approximately constant across a range of large $n$ values when diffuse priors are used. 
 
 If $\pi(\theta)$ is informative \color{black} or one wants to precisely calibrate the cumulative type I error rates of an adaptive design with interim predictive probabilities to a pre-specified $\alpha$-spending function\color{black}, Lines 7 to 13 could be implemented with sampling distribution estimates $\{\boldsymbol{\tau}(\boldsymbol{x}_{n_a, 0, r})\}_{r = 1}^R$ and $\{\boldsymbol{\tau}(\boldsymbol{x}_{n_b, 0, r})\}_{r = 1}^R$. \color{black} The resulting linear approximations for $\hat{l}_{\text{IP}, t}(\boldsymbol{x}_{n, 0, r})$ would be constructed with respect to $n$ instead of $n^2$ given the results in Appendix C.2. \color{black} These linear approximations could be used to find decision thresholds that control the type I error rate for each $n$ considered. \color{black} We provide pseudocode for this more complex implementation of Algorithm \ref{alg1} and evaluate its performance in Appendix D.3. \color{black} We evaluate the performance of Algorithm \ref{alg1} in Sections \ref{sec:ex} and \ref{sec:orvac}.

       \section{Numerical Assessment for Non-Adaptive Designs}\label{sec:ex}

       \subsection{Implications of Misspecification and Temporary Censoring}\label{sec:ex.1}

           The stylized examples in this section clarify several considerations about applying our design methodology. We first examine a collection of simple, non-adaptive designs to explore the impact of temporary censoring and misspecifying the estimating equation, $\sum_{i=1}^{n}\nabla_{\theta}\rho(\boldsymbol{X}_i, \theta) = 0$. For all designs that we consider, the structure of the time-to-event data $\boldsymbol{X}_i = (Y_i, A_i, \delta_i)$ is detailed in Section \ref{sec:methods}. In this subsection, we consider balanced randomization for the two treatment groups, and the maximum follow-up time is 5 time units, after which $Y_i$ is right censored at a value of 5.

We reconsider the loss functions $\rho_1(\boldsymbol{X}_i, \theta, \eta)$ and $\rho_2(\boldsymbol{X}_i, \theta)$ from Section \ref{sec:methods}. The first loss function corresponds to a PH model with an exponential baseline hazard, and the second is based on the partial likelihood for a PH model. All experiments aim to support the hypothesis $H_1: \theta < 0$ (i.e., $\delta_U = 0$ and $\delta_L = -\infty$), where $\theta$ is the log-hazard ratio. We assume the times for subject enrollment follow a Poisson process with mean 0.1. That is, we expect 10 subjects to enroll into the experiment per unit of time. 

In the first two of six designs that we explore, $\mathcal{F}_1$ is a degenerate data-generation family corresponding to a PH model with log-hazard ratio $-0.8$ and an $\text{EXP}(0.4)$ baseline hazard. The data are analyzed using $\rho_1(\boldsymbol{X}_i, \theta, \eta)$. Since the model is correctly specified, $\theta^*_r = -0.8$ for all simulation repetitions $r$. Design 1 uses a decision rule based on the posterior probability $\tau_{\text{PP}}(\boldsymbol{x}_n) = \Pr(H_1~|~\boldsymbol{x}_n)$. In design 2, the outcomes for all subjects who have not yet experienced the event or reached the end of follow up when the $n^{\text{th}}$ subject enrolls are temporarily right censored. Decisions are based on the ``interim'' predictive probability $\tau_{\text{IP}}(\boldsymbol{x}_n) = \Pr(\tau_{\text{PP}}(\tilde{\boldsymbol{x}}_n) \ge \gamma ~|~\boldsymbol{x}_n)$. While unconventional for a non-adaptive design, this decision procedure allows us to initially consider the impact of temporary censoring in a simplified context.

       In designs 3 and 4, $\mathcal{F}_1$ is degenerate and corresponds to a PH model with log-hazard ratio $-1.145$ and a $\text{WEI}(0.08, 2.25)$ baseline hazard. The loss function $\rho_1(\boldsymbol{X}_i, \theta, \eta)$ is again considered, this time corresponding to a proxy model where the log-likelihood -- and resulting estimating equation -- is misspecified. This data-generation process is such that $\theta^*_r \approx -0.8$ when all subjects are observed for a maximum follow-up period of 5 time units. Designs 5 and 6 are such that $\mathcal{F}_1$ corresponds to a PH model with log-hazard ratio $-0.8$ and a $\text{WEI}(0.08, 2.25)$ baseline hazard. We now use $\rho_2(\boldsymbol{X}_i, \theta)$, so $\theta^*_r = -0.8$ for all simulation repetitions $r$. Designs 3 and 5 are such that decisions are based on $\tau_{\text{PP}}(\boldsymbol{x}_n)$. Designs 4 and 6 implement temporary censoring as in design 2, and decisions are made using $\tau_{\text{IP}}(\boldsymbol{x}_n)$. For all designs, a $\mathcal{N}(0,10)$ prior is assigned to $\theta$, and a $\mathcal{N}(0,10)$ prior is independently assigned to $\log(\eta)$ for designs 1 to 4.

       For designs 2 and 4, $\tilde{F}_r(\tilde{x}~|~\theta_m, \eta_m)$ that defines the posterior predictive distribution is a PH model with an exponential baseline hazard, where both the log-hazard ratio $\theta_m$ and baseline parameter $\eta_m$ are jointly drawn from the generalized posterior in (\ref{eq:M.post}). To satisfy the condition on $\tilde{F}_r$ in Lemma \ref{lem2} when $\rho(\boldsymbol{X}_i, \theta)$ corresponds to a complete likelihood function, $\tilde{F}_r$ should be the model that corresponds to this (proxy) likelihood -- which may not coincide with the true data-generation process $F_r$. In design 6, $\tilde{F}_r(\tilde{x}~|~\theta_m)$ is a PH model where the log-hazard ratio $\theta_m$ is drawn from the generalized posterior and the baseline hazard is estimated nonparametrically using Breslow's estimator \citep{breslow1972contribution}. Under $\rho_2(\boldsymbol{X}_i, \theta)$, the condition on $\tilde{F}_r$ in Lemma \ref{lem2} is less restrictive; it is satisfied for \emph{any} PH model, regardless of the baseline hazard used to generate truncated predictive data from $\dot{F}$ in (\ref{eq:ppd}). Table \ref{tab:artif} overviews the data-generation processes, impact of temporary censoring, and decision rules to conclude success for all six designs.

       \begin{table}[!ht]
       \centering
\begin{tabular}{ccccc}
Design & Baseline  & $\rho$    & Temporary Censoring & Decision Rule \\ \hline
1      & EXP  & $\rho_1$   & No                   &    $\tau_{\text{PP}}(\boldsymbol{x}_n) \ge 0.99$           \\
2      & EXP    & $\rho_1$ & Yes                  &       $\Pr(\tau_{\text{PP}}(\tilde{\boldsymbol{x}}_n) \ge 0.99 ~|~\boldsymbol{x}_n) \ge 0.95$        \\
3      & WEI & $\rho_1$ & No                   &        $\tau_{\text{PP}}(\boldsymbol{x}_n) \ge 0.99$       \\
4      & WEI & $\rho_1$ & Yes                   &      $\Pr(\tau_{\text{PP}}(\tilde{\boldsymbol{x}}_n) \ge 0.99 ~|~\boldsymbol{x}_n) \ge 0.95$  \\
5      & WEI & $\rho_2$ & No                   &      $\tau_{\text{PP}}(\boldsymbol{x}_n) \ge 0.99$   \\
6      & WEI & $\rho_2$ & Yes                   &      $\Pr(\tau_{\text{PP}}(\tilde{\boldsymbol{x}}_n) \ge 0.95 ~|~\boldsymbol{x}_n) \ge 0.80$
\end{tabular}
\caption{Summary of the baseline hazard, loss function, temporary censoring, and decision rules to conclude success for all designs considered.}\label{tab:artif}
\end{table}

\begin{figure}[!tb]
      \centering
		\includegraphics[width = 0.85\textwidth]{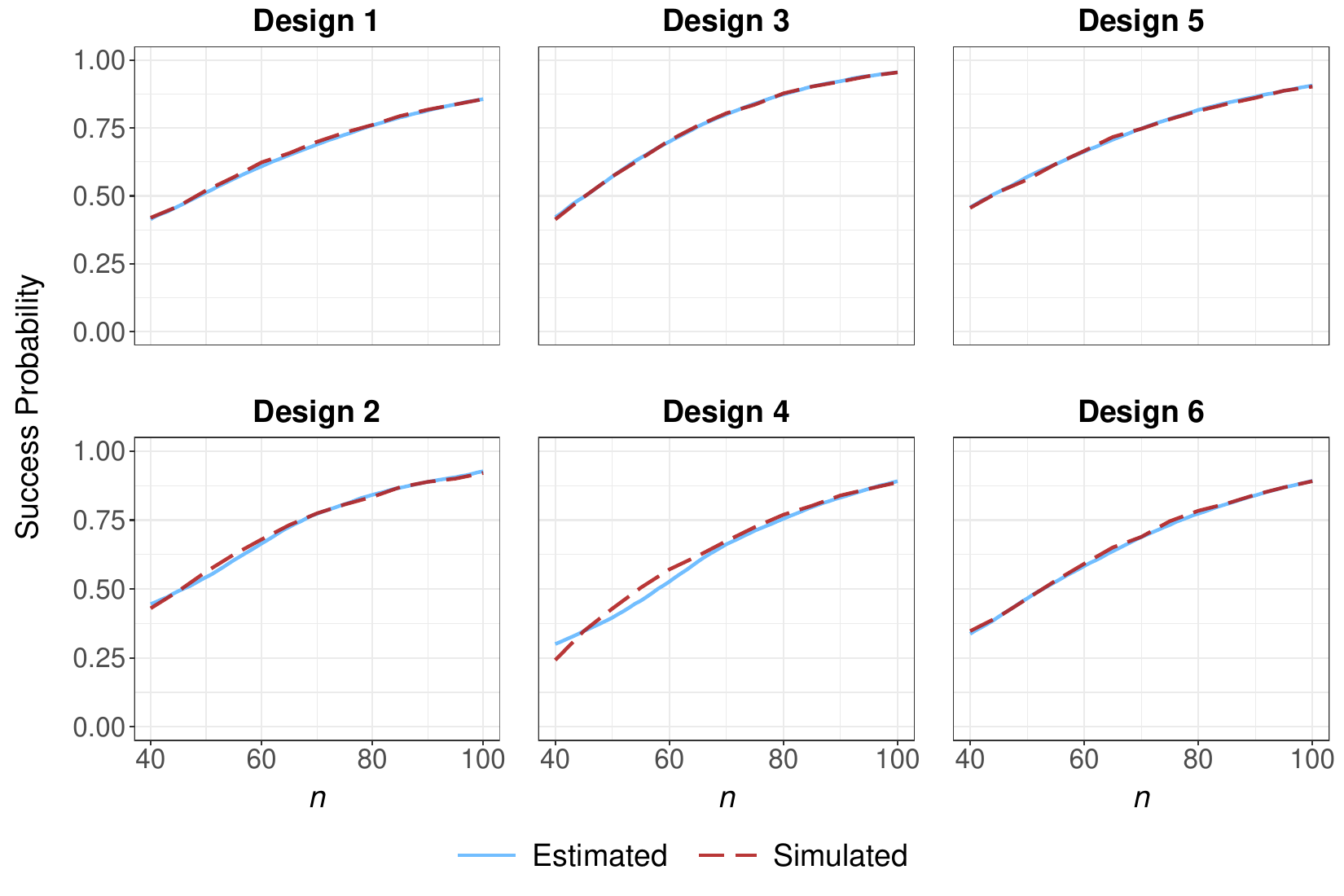} 

		\caption{\label{fig:artif} Power curves obtained using Algorithm \ref{alg1} (estimated) or naive simulation (simulated) for the six simple designs.} 
	\end{figure}

Instead of being selected by Algorithm \ref{alg1} to control the type I error rate, the decision thresholds in Table \ref{tab:artif} were chosen so that the probability of concluding success (i.e., power) for each design varies widely as a non-linear function of $n$ across the sample sizes we consider. Control of the type I error rate will be considered in Section \ref{sec:ex.2}, and this example broadly aims to assess how well Algorithm \ref{alg1} estimates power independent of sample size determination. For all designs, Algorithm \ref{alg1} was implemented with $R = 10^4$ and $M = 10^3$ repetitions and sample sizes of $n_a = 45$ and $n_b = 95$. Figure \ref{fig:artif} visualizes the power curve with respect to $n$ for all designs. The blue curves were \emph{estimated} using linear approximations to logits of posterior summaries at only two sample sizes ($n_a$ and $n_b$) using the process in Lines 4 to 13 of Algorithm \ref{alg1}. The red curves were \emph{simulated} by independently generating samples $\boldsymbol{x}_n$ to estimate sampling distributions of $\boldsymbol{\tau}(\boldsymbol{x}_{n})$ at $n = \{40, 45, \dots, 100 \}$.

We use the red curves as surrogates for the true power;
although they are impacted by simulation variability, this impact is minimal with $R = 10^4$ and $M = 10^3$ repetitions for this example. Apart from design 4, we observe good alignment between the blue and red curves. For those five designs, the maximum absolute difference between the design-specific blue and red power curves ranges between 0.0086 and 0.0171.  However, the two curves diverge for design 4: the linear approximations from Algorithm \ref{alg1} do a poor job of approximating trends in the true sampling distribution of posterior summaries.  The maximum absolute difference between those design-specific blue and red power curves is 0.0582.  This divergence is not surprising. The temporary censoring in design 4 is such that the true value of $\theta^*_r$ \emph{changes} as the temporary censoring proportion decreases with $n$ (see Appendix E.1 of the supplement for further discussion). 

       In that case, the result in (\ref{eq:joint.M}) does not hold true, and our methodology in Algorithm \ref{alg1} cannot be applied. The results for designs 2 and 6 illustrate that we can apply our methodology in the presence of temporary censoring when the estimating equation resulting from $\rho(\boldsymbol{X}, \theta)$ holds. Design 3 illustrates that we can also apply our approach when the estimating equation is misspecified in a design without temporary censoring. Design 6 demonstrates that we may be able to circumvent issues with misspecification entirely by leveraging a flexible and robust loss function. Nevertheless, the combination of loss-function misspecification \emph{and} temporary censoring does present issues for our proposed methods (hence the condition on $\sum_{i=1}^{n}\nabla_{\theta}\rho(\boldsymbol{X}_i, \theta) = 0$ in Lemma \ref{lem1}).

       The good performance of our method for all designs except design 4 is expected because the conditions for Lemmas \ref{lem1} to \ref{lem3} are satisfied. The results in (\ref{eq:M.BvM}) and (\ref{eq:joint.M}) hold true for designs 1 to 3 with $\rho_1(\boldsymbol{X}_i, \theta, \eta)$ by \citet{huber2009robust}: the loss function corresponds to a likelihood where $\theta^*_r$ (and $\eta^*_r$) are in the interior of the parameter space, the maximum follow-up time is independent of the survival time, and the only covariate $A_i \in \{0,1\}$ in the PH model is bounded. For designs 5 and 6, the results in (\ref{eq:M.BvM}) and (\ref{eq:joint.M}) hold true by combining similar logic with counting process theory from \citet{fleming2005counting}.

All simulations in this paper were run on a high-performance computing server with 72 cores. Even when using the Laplace approximation within Algorithm \ref{alg1}, it took roughly 3 minutes to estimate each blue curve for designs 1 to 4 in Figure \ref{fig:artif}; 15 minutes were required for designs 5 and 6. It took roughly 20 minutes to obtain each red curve for designs 1 to 4 using naive simulation (100 minutes for designs 5 and 6). Our methodology is much more computationally efficient because we need only estimate the sampling distribution of posterior summaries at two values of $n$. 

In Appendix E.1, we conduct additional simulations that repurpose designs from this subsection. Those numerical studies illustrate how to assess the performance of Algorithm \ref{alg1}  by estimating the sampling distribution of $\boldsymbol{\tau}(\boldsymbol{x}_n)$ at the sample size recommendation $n_c$, and \color{black} we show how this third sampling distribution estimate could be used to improve the performance of Algorithm \ref{alg1} if $n_c$ is not near $n_a$ or $n_b$. \color{black} Moreover, these simulations explore various choices for the number of repetitions $R$ and $M$ as well as the sample sizes $n_a$ and $n_b$ in Algorithm \ref{alg1}. \color{black} Based on those simulations, we recommend (i) using $R=10^4$ and $M = 10^3$ where computationally feasible and (ii) enforcing a minimum difference between $n_a$ and $n_b$ (i.e., a 20\% difference on the relative scale) to limit the impact of Monte Carlo error. \color{black} An example design with an equivalence-based hypothesis (i.e., when $\delta_L$ and $\delta_U$ are both finite) is also considered.

       \subsection{Control of the Type I Error Rate}\label{sec:ex.2}

       Here, we explore a stylized example that illustrates several considerations for control over the type I error rate. In this example, we design a study to assess the effectiveness of a 12-week, fitness-based intervention. This intervention is aimed at reducing HOMA-IR (Homeostasis Model Assessment of Insulin Resistance) in individuals with prediabetes \citep{amaravadi2024effectiveness}. The target of inference $\theta$ is the median percentage decrease in HOMA-IR after completing the intervention. The corresponding loss function is $\rho(x_i, \theta) = \lvert x_i - \theta \rvert$, where $x_i$ is the percentage decrease for participant $i = 1, \dots, n$.  This experiment aims to demonstrate a material benefit of the intervention by supporting the hypothesis $H_1: \theta > 1.1$ (i.e., $\delta_L = 1.1$ and $\delta_U = \infty$). 

       We now overview the data-generation families under which power and the type I error rate are defined, and full details on data generation are provided in Appendix E.2 of the supplement. We define power for this example under the predictive approach: $\mathcal{F}_1$ is such that the true median $\theta^*_r$ follows a $\mathcal{N}(1.2, 0.075^2)$ distribution that is truncated between 1.15 and 1.25. We consider type I error under the conditional approach such that $\theta^*_r = 1.1$ in all simulation repetitions. For all simulation repetitions, $F_r$ from $\mathcal{F}_1$ or $\mathcal{F}_0$ corresponds to a mixture gamma distribution. We define a target power of $1 - \beta = 0.9$ and type I error rate of $\alpha = 0.05$.  
       The prior $\pi(\theta)$ for this example is $\text{UNIF}(0,10)$. 
       
       To control the type I error rate, we allow the learning rate $\omega$ in (\ref{eq:M.post}) to differ from 1. \color{black} This consideration is particularly relevant when the loss function is not based on a correctly specified (partial) likelihood. \color{black} In the following discussion, we let $A_r = \mathbb{E}_{F_r}\left[ \nabla_{\theta}^2 \rho(\boldsymbol{X},\theta) \right] \big|_{\theta = \theta^*_r}$ and $B_r = \mathbb{E}_{F_r}\left[ \nabla_{\theta} \rho(\boldsymbol{X},\theta)^2\right]\big|_{\theta = \theta^*_r}$. Based on a Taylor expansion of the log-generalized posterior, the limiting \color{black} unscaled \color{black} variance in (\ref{eq:M.BvM}) is $\sigma^2_r = (\omega A_r)^{-1}$. For this example, the result in (\ref{eq:joint.M}) holds true by \citet{huber2009robust}; 
       the limiting \color{black} unscaled \color{black} variance of the M-estimator $\hat{\theta}^{_{(n)}}_r$ is thus $B_r/A_r^2$. The coverage of credible sets -- and the corresponding type I error rate -- is asymptotically calibrated when the limiting variances of the M-estimator and generalized posterior coincide \citep{chernozhukov2003mcmc}. These variances coincide for scalar $\theta$ when $\omega = A_r/B_r$. For this example, we show that this value of $\omega$ is $2f(\theta^*_r)$ in Appendix E.2. 
       
       Our theory in Section \ref{sec:proxy} treats $\omega$ as a constant. For this example, we estimate $\omega = 2f(\theta^*_r)$ from the data by obtaining a nonparametric kernel density estimate of the percentage-decrease distribution and evaluating this density estimate at the sample median. More sophisticated approaches exist to calibrate the credible intervals of generalized posteriors (see e.g., \citet{syring2019calibrating}), but we use this pragmatic approach because (i) it must be applied across many simulation repetitions and (ii) we can further calibrate the type I error rate by tuning the decision thresholds. Our methods in Section \ref{sec:power} can still be applied to this example when estimating $\omega$ using this pragmatic approach. As $n \rightarrow \infty$, the estimated value for $\omega$ converges to $2f(\theta^*_r)$.  

                   \begin{figure}[!tb]
      \centering
		\includegraphics[width = 0.825\textwidth]{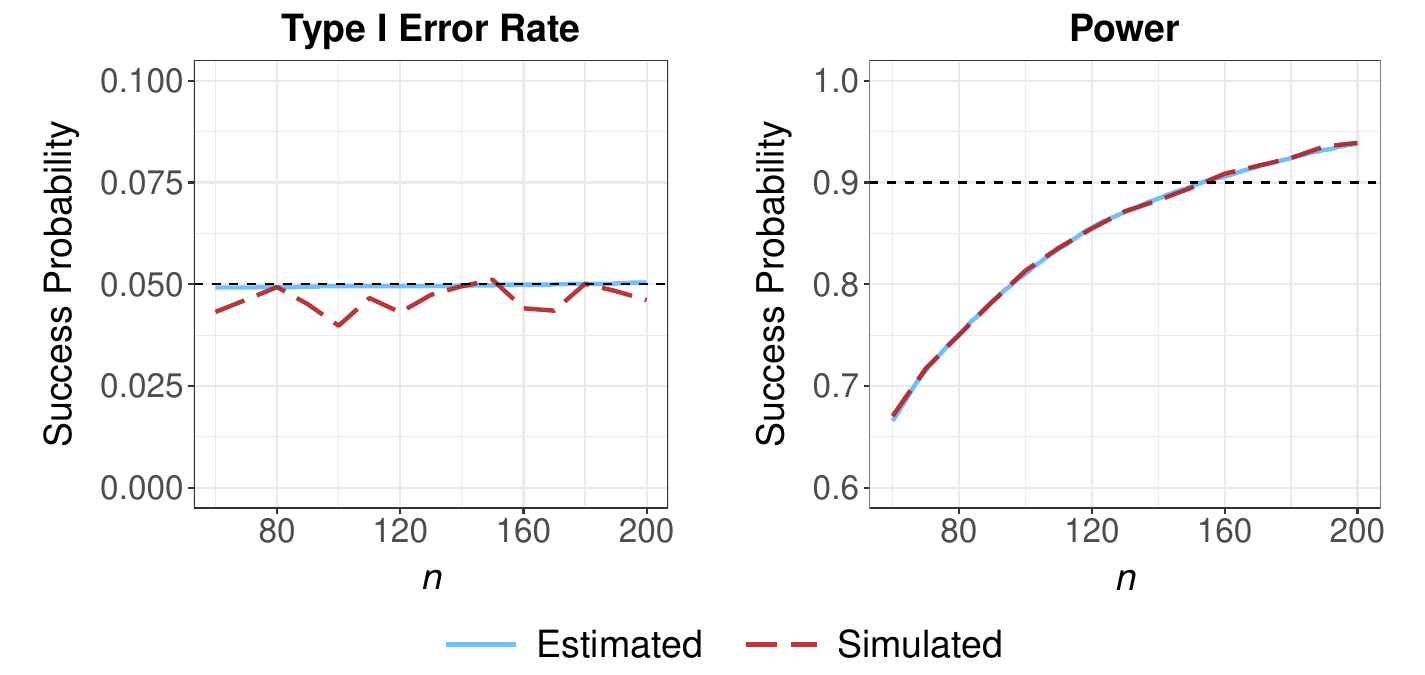} 

		\caption{\label{fig:med} The type I error rate and power of the experiment as a function of $n$, obtained using Algorithm \ref{alg1} (estimated) and naive simulation (simulated). The dotted horizontal lines denote $\alpha = 0.05$ and $1 - \beta = 0.9$.} 
	\end{figure}

       When implementing Algorithm \ref{alg1} with $R = 10^4$, an initial sample size of $n_a = 180$ was selected. In Line 3 of Algorithm \ref{alg1}, we chose $\gamma = 0.9425$ as the smallest threshold that bounded the type I error rate by $\alpha = 0.05$. For the selected family $\mathcal{F}_1$ and $n_a = 180$, the estimated power was $0.9240 > 1-\beta$. We next explored a sample size of $n_b = 80$ in Line 6 to facilitate assessment of the operating characteristics for $n \in [60, 200]$. Under the predictive approach, the logits of posterior probabilities were split into 10 subgroups based on the order statistics of their $\theta^*_r$ values before constructing the linear approximations. We also estimated the sampling distribution of $\tau_{\text{PP}}(\boldsymbol{x}_n)$ at $n = \{60, 70, \dots, 200\}$ to assess the performance of Algorithm \ref{alg1}. For this example, we estimated these sampling distributions -- and constructed linear approximations using $n_a = 180$ and $n_b = 80$ -- under both $\mathcal{F}_0$ and $\mathcal{F}_1$ to consider power and the type I error rate as functions of $n$. These results are visualized in Figure \ref{fig:med}.

    Figure \ref{fig:med} confirms the good alignment between the results produced using our linear approximations and those obtained using naive simulation.   The maximum absolute difference between the blue and red curves is 0.0097 for the type I error rate and 0.0047 for power.  Apart from the impact of simulation variability, the type I error rate is roughly constant as $n$ changes, hence why we do not typically construct linear approximations under $\mathcal{F}_0$ in Algorithm \ref{alg1}. We approximated all generalized posteriors in this subsection using Markov chain Monte Carlo (MCMC) with 2 chains consisting of $10^3$ burn-in iterations and $10^4$ retained draws per chain. It took 3 and 22 minutes, respectively, to obtain each blue and red curve in Figure \ref{fig:med} with these MCMC settings, illustrating the computational benefit of our methodology. 

    \section{Redesign of an Adaptive Clinical Trial}\label{sec:orvac}

The Optimizing Rotavirus Vaccine in Aboriginal Children
(ORVAC) trial was a Bayesian adaptive clinical trial to test the effectiveness of a third dose of Rotarix rotavirus vaccine in Australian Indigenous infants in providing improved protection against gastroenteritis \citep{middleton2019orvac, jones2020orvac}. The third dose was considered as the active treatment, with the usual two doses considered as a placebo. In this section, we showcase our computationally efficient methodology by redesigning this trial. One of ORVAC's primary outcomes was the time from randomization to medical attendance for which the primary reason for presentation was acute gastroenteritis or diarrhea illness. For the purposes of sample size determination, we focus on this outcome where the target of inference $\theta$ is the log-hazard ratio between active and placebo groups. The two treatments were randomized in a ratio of 1:1 throughout the trial.

We suppose children between 6 and 12 months of age enroll in the trial, and all children have a maximum follow-up period of 30 months. In the actual trial, the maximum sample size was 1000 participants. The first interim analysis occurred after 250 participants were enrolled, with subsequent analyses after every 50 enrollments. We retain the $T=16$ potential analyses and constants $(c_1, c_2, \dots, c_{16}) = (1, 1.2, \dots, 4)$ from ORVAC for this illustration. Moreover, we suppose that participant enrollment times follow a Poisson process with mean $0.06$. That is, we expect 50 participants to enroll during each 3-month period.

Trial design for ORVAC was considered using PH models with Weibull and exponential distributions for the baseline hazard. In this section, we use the robust loss function $\rho_2(\boldsymbol{X}_i, \theta)$ resulting from the partial likelihood in (\ref{eq:partial}) with a learning rate of $\omega = 1$. We consider a $\mathcal{N}(0, 100)$ prior for $\theta$ and use the Laplace approximation to obtain all posterior summaries. The data $\boldsymbol{X}_i = (Y_i, A_i, \delta_i)$ for participant $i = 1, \dots, n$ follow the structure from Section \ref{sec:methods}, where the time $Y_i$ is measured in months for this example. As in Section \ref{sec:ex.1}, we aim to support the hypothesis $H_1: \theta < 0$, and the results in (\ref{eq:M.BvM}) and (\ref{eq:joint.M}) \color{black}now hold true by \citet{sellke1983sequential}.  \color{black}

The degenerate data-generation family $\mathcal{F}_1$ under which we consider power is a PH model with log-hazard ratio $-0.2$ and the baseline hazard is characterized by a five-knot spline. The family  $\mathcal{F}_0$ under which we consider type I error modifies $\mathcal{F}_1$ by setting the log-hazard ratio to 0. The data-generation process $\tilde{F}_r(\tilde{x}~|~\theta_m)$ that defines $\dot{F}$ in (\ref{eq:ppd}) is a PH model where the log-hazard ratio $\theta_m$ is drawn from the generalized posterior and the baseline hazard is estimated using standard Bayesian inference via a spline with five knots. This choice for $\tilde{F}_r(\tilde{x}~|~\theta_m)$ is distinct from that considered in Section \ref{sec:ex.1}, but it also satisfies  the condition on $\tilde{F}_r$ in Lemma \ref{lem2}. 

        \begin{figure}[!tb]
      \centering
		\includegraphics[width = \textwidth]{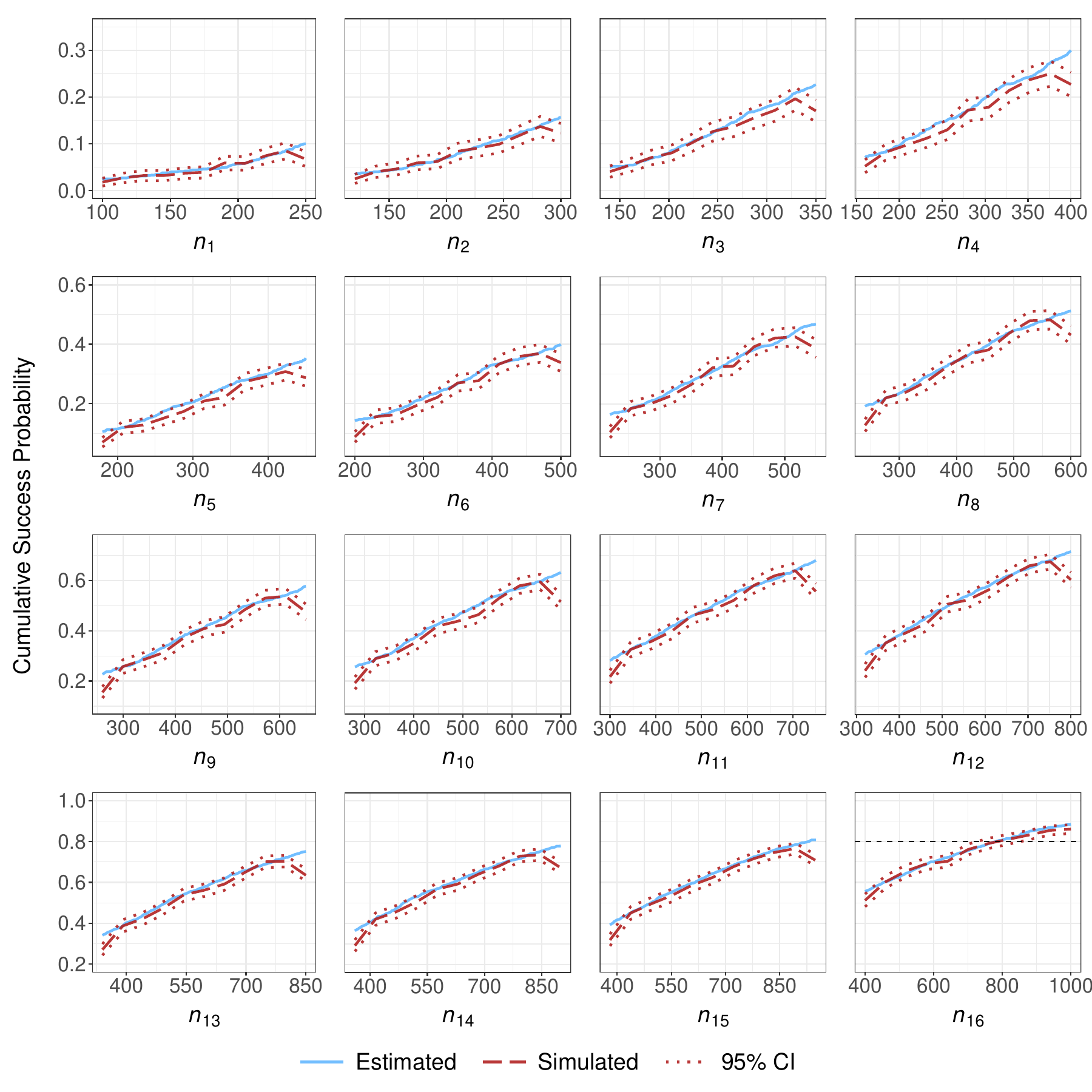} 

		\caption{\label{fig:suc} Cumulative probability of stopping for success at analysis $t$ as a function of $n_t$ obtained using Algorithm \ref{alg1} (estimated) and naive simulation (simulated). Pointwise 95\% confidence intervals are based on naive simulation. The dashed black line represents the target power of $1 - \beta = 0.8$.} 
	\end{figure}

This design has success thresholds $\boldsymbol{\gamma}$ and $\boldsymbol{\xi}$ along with failure thresholds $\boldsymbol{\kappa}$. At the interim analyses ($t = 1, \dots, 15$), we stop for success if $\tau_{\text{IP}, t}(\boldsymbol{x}_n) \ge \xi_t$. If the experiment did not stop for success, we then stop for failure if $\tau_{\text{FP}, t}(\boldsymbol{x}_n) < \kappa_t$. At the final analysis $(t = 16)$, we stop for success if $\tau_{\text{PP}, T}(\boldsymbol{x}_n) \ge \gamma_T$. We aim to obtain a sample size $n$ and decision thresholds that attain a target power of $1 - \beta = 0.8$ under $\mathcal{F}_1$ and bound the family-wise error rate (FWER) by $\alpha = 0.05$ under $\mathcal{F}_0$.

We implemented Algorithm \ref{alg1} with $R = 10^3$ and $M = 500$ repetitions as well as $n_a = 115$ and $n_b = 235$ to explore a similar range of sample sizes as those considered in ORVAC. Based on the sampling distribution estimate of $\boldsymbol{\tau}(\boldsymbol{x}_{n})$ at $n_a = 115$, we considered several sets of potential decision thresholds and ultimately selected thresholds $\boldsymbol{\gamma} = 0.975\times \mathbf{1}_{16}$, $\boldsymbol{\kappa} =  0.05\times \mathbf{1}_{15}$, and $\boldsymbol{\xi} = (0.97, 0.965, \dots, 0.9)$ that maintained the desired FWER. Figure \ref{fig:suc} visualizes the cumulative probability of stopping for success at all analyses as a function of $n_t$. The blue curves were estimated using linear approximations from Algorithm \ref{alg1}. The dashed red curves were simulated by generating samples $\boldsymbol{x}_n$ to estimate sampling distributions of $\boldsymbol{\tau}(\boldsymbol{x}_{n})$ at $n = \{100, 115, \dots, 250 \}$. The dotted red curves represent pointwise 95\% confidence intervals for the cumulative success probability obtained using a variance estimate of $R^{-1}\hat{p}(1 - \hat{p})$, where $\hat{p}$ represents the estimated cumulative probability.

While Figure \ref{fig:suc} illustrates relatively good alignment between the blue and red curves, there is a substantial impact of simulation variability. Since $\mathcal{F}_1$ is such that $H_1$ is true, the red curves should increase with $n_t$ in theory, but the cumulative success probability based on $n = 250$ is smaller than that calculated at $n = 235$ in many subplots. Simulation variability is substantial because only $R = 10^3$ and $M = 500$ repetitions were used. For this example, it was not computationally feasible to use $R = 10^4$ and $M = 10^3$ repetitions as recommended in Appendix E.1. Even with these smaller numbers of repetitions, it took 30 hours using parallelization with 72 cores to get the set of red curves in Figure \ref{fig:suc}; it took only 4 hours to obtain the blue curves using Algorithm \ref{alg1}. 

The pointwise confidence intervals in Figure \ref{fig:suc} quantify the variability associated with data that could be observed under $\mathcal{F}_1$. However, these confidence intervals do not account for the additional Monte Carlo error associated with estimating $\tau_{\text{IP}, t}(\boldsymbol{x}_n)$ or $\tau_{\text{FP}, t}(\boldsymbol{x}_n)$ for a given sample $\boldsymbol{x}_n$ using only $M = 500$ repetitions. These confidence intervals are thus anti-conservative. 
The good performance of Algorithm \ref{alg1} for the similar but simpler designs 5 and 6 in Section \ref{sec:ex.1} suggests there is more value in obtaining a smaller number of precise sampling distribution estimates than a larger number of noisy estimates. Our proposed methodology helpfully provides a framework to implement experimental design using two such estimates of $\boldsymbol{\tau}(\boldsymbol{x}_{n})$. 

A plot similar to Figure \ref{fig:suc} for the cumulative probability of failure across analyses is provided in Appendix F of the supplement. This plot indicates good agreement between the results obtained using Algorithm \ref{alg1} and naive simulation. For this example, the recommended sample size for the first interim analysis was $n_1 = 195$, which translates to a maximum sample size of $n_{16} = 780$.  At this sample size, we estimated the sampling distribution of $\boldsymbol{\tau}(\boldsymbol{x}_n)$ to acquire confirmatory estimates of power (0.799) and the FWER (0.043), \color{black}also based on $R = 10^3$ and $M = 500$ repetitions. \color{black}These confirmatory estimates are close to the target power and desired FWER, corroborating the suitable performance of our computationally efficient approach. 

 \section{Discussion}\label{sec:disc}

 In this paper, we proposed an efficient framework to design Bayesian experiments based on generalized posteriors under the hybrid approach. This framework determines the minimum sample size required to attain desired power for the study while bounding the type I error rate. While we emphasized adaptive designs with temporary censoring for time-to-event outcomes, our methodology can be broadly applied in various settings where the M-estimator and generalized posterior are approximately normal. The proposed methods facilitate robust designs that do not rely on the correct specification of a likelihood function. The computational efficiency of our framework is motivated by considering a proxy for the joint sampling distribution of various posterior and posterior predictive probabilities based on generalized posteriors. We use the behaviour in this large-sample proxy distribution to justify estimating true sampling distributions at only two sample sizes. Our method therefore substantially reduces the number of simulation repetitions required to design Bayesian experiments under the hybrid approach. 

 The methodology in this paper could be extended to accommodate more complex adaptive designs based on generalized posteriors. In this article, our approach requires that practitioners fix treatment allocations in multi-arm designs prior to collecting data. This constraint precludes group sequential designs that leverage response-adaptive randomization, an increasingly prominent feature in Bayesian clinical trials.  Under response-adaptive randomization, the result in (\ref{eq:joint.M}) does not hold true: the entries in the matrix $\boldsymbol{V}_r$ for later stages depend on the (dynamic) treatment allocation in previous stages. The sample size $n$ therefore cannot be factored out of the covariance matrix in (\ref{eq:joint.M}), which is a requirement for the asymptotic linearity in Theorem \ref{thm1}. If the treatment allocation becomes more (less) statistically efficient in the next stage, the logits of the posterior summaries generally exhibit strict convexity (concavity) that does not dissipate as $n \rightarrow \infty$.  

 Finally, we acknowledge that the Bayesian-frequentist hybrid approach to design may not be ideal for certain studies. Practitioners may instead prefer a more Bayesian design approach in which decision making could account for enrollment costs, posterior precision criteria, or the success probability for a future confirmatory experiment (see e.g., \citet{pericchi2016adaptive,calderazzo2022decision,hagar2023economical}). Future research could consider how to efficiently implement extensions that accommodate response-adaptive randomization or design criteria beyond frequentist operating characteristics by developing more complex proxy sampling distributions.

 \section*{Supplementary Material}
These materials include the proofs of Lemmas \ref{lem1} to  \ref{lem3} and Theorem \ref{thm1}, as well as an analogue to Algorithm \ref{alg1} for non-adaptive designs and additional content for the examples in Sections \ref{sec:ex} and \ref{sec:orvac}. The code to conduct the numerical studies in the paper is available online: \url{https://github.com/lmhagar/GeneralizedSSD}.


	\section*{Funding}
 
LH was supported by The University of Queensland's Health Research Accelerator initiative. JM was supported by an Australian Research Council Discovery Project (DP200101263).
	

\bibliographystyle{chicago}


\end{document}